\documentclass[prd, twocolumn, nofootinbib, floatfix]{revtex4-1}
\usepackage{silence}
\usepackage{lmodern} 
\usepackage[T1]{fontenc}
\usepackage{textcomp}
\usepackage[scr=boondoxo,frak=boondox,bb=boondox]{mathalfa}
\pdfoutput=1
\usepackage{makecell, multirow}
\setcellgapes{3pt}
\usepackage{hyperref}
\usepackage{amsmath}
\usepackage{tabularx}
\usepackage[vlines]{tabularht}
\usepackage{adjustbox}
\usepackage{tikz}
\usepackage[]{draftfigure}
\usepackage{amssymb}
\usepackage{graphicx}
\usepackage{float} 
\usepackage{color}
\usepackage{color,soul}
\usepackage{ulem}
\usepackage{multirow}
\usepackage{epsfig}
\usepackage{epstopdf}
\usepackage{rotating}
\usepackage{afterpage}
\usepackage{soul}
\usepackage{fancyvrb}
\usepackage{lineno}
\usepackage{amsmath}
\usepackage{xspace}
\usepackage{hyperref}
\usepackage{url}
\usepackage{amssymb,latexsym,mathrsfs}
\usepackage{multirow}
\usepackage{bigints}
\usepackage{soul}
\usepackage{graphicx}
\usepackage{subcaption}
\usepackage{dcolumn}
\usepackage{bm}
\usepackage{epsfig}
\usepackage{caption}
\usepackage{color}
\usepackage{multirow}
\usepackage{makecell}
\usepackage{textcomp}
\usepackage[title]{appendix}
\usepackage{tabularx}
\usepackage{natbib}
\setcitestyle{authoryear,open={(},close={)}} 

\hypersetup{
    colorlinks=true,
    linkcolor=red,
    citecolor=blue,
} 

\newcommand{\be}{\begin{equation}}
\newcommand{\ee}{\end{equation}}
\newcommand{\bea}{\begin{eqnarray}}
\newcommand{\eea}{\end{eqnarray}}
 
\definecolor{mediumpurple}{rgb}{0.58, 0.44, 0.86}

\begin{document}
\title{ Using Host Galaxy Photometric Redshifts to Improve Cosmological
 Constraints with Type Ia Supernova in the LSST Era
}

\newcommand{\URLSNANA}{\url{https://github.com/RickKessler/SNANA}}
\newcommand{\URLLSST}{\url{www.lsst.org}}
\newcommand{\URLDESC}{\url{https://lsstdesc.org}}
\newcommand{\URLOPSIM}{\url{http://opsim.lsst.org/runs/minion_1016/data/minion_1016_sqlite.db.gz}}
\newcommand{\HOSTLIB}{{\tt HOSTLIB}}
\newcommand{\SALTII}{{\sc SALT-II}}
\newcommand{\DESSN}{DES-SN}
\newcommand{\lowz}{low-$z$~}
\newcommand{\mosfit}{{\tt MOSFiT}\xspace}
\newcommand{\bands}{$ugrizy$}
\newcommand{\PLASTICC}{Photometric LSST Astronomical Time Series Classification Challenge}
\newcommand{\acro}{{\tt PLAsTiCC}}
\newcommand{\SNANA}{{\tt SNANA}}
\newcommand{\pippin}{{\tt Pippin}}
\newcommand{\LSST}{Large Synoptic Survey Telescope}
\newcommand{\DES}{Dark Energy Survey}
\newcommand{\OPSIM}{\verb|OpSim|}
\newcommand{\Spec}{Spectroscopic}
\newcommand{\spec}{spectroscopic}
\newcommand{\specy}{spectroscopically}
\newcommand{\ZPHOT}{{\bf\tt ZPHOT}}
\newcommand{\zSpec}{z_{\rm spec}}
\newcommand{\LCDM}{\Lambda{\rm CDM}}
\newcommand{\OL}{\Omega_{\Lambda}}
\newcommand{\OM}{\Omega_{\rm M}}
\newcommand{\zcmb}{z_{\rm cmb,true}}
\newcommand{\RateUnit}{{\rm yr}^{-1}{\rm Mpc}^{-3}}
\newcommand{\Ftrue}{F_{\rm true}}
\newcommand{\sigF}{\sigma_{F}}
\newcommand{\sigFtrue}{\sigma_{\rm Ftrue}}
\newcommand{\zSN}{z_{\rm SN}}
\newcommand{\zHOST}{z_{\rm HOST}}
\newcommand{\Ngen}{$N_{\rm gen}$} %

\newcommand{\NSAMPLE}{25} 
\newcommand{\NBIASCORTOT}{$3.1 \times 10^6$}
\newcommand{\NBIASCORHIZ}{$2.6 \times 10^6$}
\newcommand{\NBIASCORLOZ}{$4.4 \times 10^5$}
\newcommand{\NSYST}{7}
\newcommand{\dz}{\delta z}
\newcommand{\fout}{f_{\rm out}}
\newcommand{\sigIQR}{\sigma_{\rm IQR}}
\newcommand{\ztrue}{z_{\rm true}}
\newcommand{\zphot}{z_{\rm phot}}
\newcommand{\zspec}{z_{\rm spec}}
\newcommand{\zcheat}{z_{\rm cheat}}
\newcommand{\Dz}{\Delta z_{(1+z)}}
\newcommand{\Pfit}{P_{\rm fit}}

\newcommand{\G}{G_{\rm host}}
\newcommand{\mubias}{\Delta\mu_{\rm bias}}
\newcommand{\mutrue}{\mu_{\rm true}}

\newcommand{\dzsyst}{\Delta z_{\rm syst-z}}
\newcommand{\dmusyst}{\Delta\mu_{\rm syst-z}}

\newcommand{\alphaTrueSym}{\alpha_{\rm true}}
\newcommand{\betaTrueSym}{\beta_{\rm true}}
\newcommand{\alphaTrueVal}{0.14}
\newcommand{\betaTrueVal}{3.1}

\newcommand{\sigmu}{\sigma_{\mu}}
\newcommand{\sigmubar}{\overline{\sigma_{\mu}}}

\newcommand{\sigz}{\sigma_{z}}
\newcommand{\sigR}{\sigma_{R}}
\newcommand{\sigint}{\sigma_{\rm int}}

\newcommand{\COVsyst}{{\rm COV}_{\rm syst}}
\newcommand{\COVsysti}{{\rm COV}_{{\rm syst},i}}
\newcommand{\COVstat}{{\rm COV}_{\rm stat}}

\newcommand{\wCDM}{$w$CDM}
\newcommand{\wwCDM}{$w_0w_a$CDM}
\newcommand{\ww}{$w_0$-$w_a$}

\newcommand{\NZBIN}{\textcolor{red}{14}}

\newcommand{\AVGwbias}{\langle w$-bias$\rangle}
\newcommand{\AVGwsigbiaszspecsyst}{$0.025$}
\newcommand{\AVGwsigbiaszphotsyst}{$0.023$}
\newcommand{\AVGwsig}{\langle\sigma_w\rangle}
\newcommand{\STDw}{{\rm STD}_w}

\newcommand{\AVGwwbias}{\langle {w_0}$-bias$\rangle}
\newcommand{\AVGwwsig}{\langle\sigma_{w_0}\rangle}
\newcommand{\STDww}{{\rm STD}_{w_0}}

\newcommand{\AVGwabias}{\langle {w_a}$-bias$\rangle}
\newcommand{\AVGwasig}{\langle\sigma_{w_a}\rangle}
\newcommand{\STDwa}{{\rm STD}_{w_a}}

\newcommand{\AVGFoM}{{\langle}{\rm FoM}{\rangle}}

\newcommand{\RatioFoM}{{\cal R}_{{\rm FoM},i}}
\newcommand{\FoM}{{\rm FoM}}
\newcommand{\FoMStat}{{\rm FoM}_{\rm stat}}
\newcommand{\FoMSysti}{{\rm FoM}_{{\rm syst},i}}
\newcommand{\AVGFOMzspecstat}{$136$}
\newcommand{\AVGFOMzspecsyst}{$95$}
\newcommand{\AVGFOMzphotstat}{$237$}
\newcommand{\AVGFOMzphotsyst}{$145$}
\newcommand{\AVGwbiaszphotsyst}{$0.0091$}
\newcommand{\AVGwabiaszphotsyst}{$0.0363$}
\newcommand{\AVGwbiaszspecsyst}{$0.0140$}
\newcommand{\AVGwabiaszspecsyst}{$0.0662$}

\author{Ayan Mitra${}^{1,2,3}$, Richard Kessler${}^{4,5}$, Surhud More$^{1}$, Renee Hlozek$^{6}$ AND \\ The LSST Dark Energy Science Collaboration \\ }  
\affiliation{
${}^1$The Inter-University Centre for Astronomy and Astrophysics (IUCAA), Post Bag 4, Ganeshkhind, Pune 411007, India \\
${}^2$Energetic Cosmos Laboratory, Nazarbayev University, Nur-Sultan 010000, Kazakhstan \\
${}^3$Kazakh-British Technical University, Almaty, Kazakhstan \\
${}^4$Kavli Institute for Cosmological Physics,~University of Chicago,~Chicago,~IL 60637,~USA \\
${}^5$Department of Astronomy and Astrophysics,~University of Chicago,~5640 South Ellis Avenue,~Chicago, IL 60637,~USA \\
${}^6$University of Toronto, 27 King's College Cir, Toronto, ON M5S, Canada \\
}




\date{\today} 

\begin{abstract}
We perform a rigorous cosmology analysis on simulated type Ia supernovae (SN~Ia) and evaluate the
improvement from including photometric host-galaxy redshifts compared to using 
only the ``$\zspec$'' subset with \spec\ redshifts from the host or SN.
We use the Deep Drilling Fields (${\sim}50$~deg$^2$) from the 
Photometric LSST Astronomical Time-Series Classification Challenge (\acro), 
in combination with a low-$z$ sample based on Data Challenge2 (DC2).
The analysis includes 
light curve fitting to standardize the SN brightness,
a high-statistics simulation to obtain a bias-corrected Hubble diagram,
a statistical+systematics covariance matrix including calibration and photo-$z$ uncertainties, 
and
cosmology fitting with a prior from the cosmic microwave background.
Compared to using the $\zspec$ subset,
including events with SN+host photo-$z$ results in 
i) more precise distances for $z>0.5$, 
ii) a Hubble diagram that extends 0.3 further in redshift, and
iii) a 50\% increase in the Dark Energy Task Force figure of merit (FoM) based
on the \wwCDM\ model.
Analyzing \NSAMPLE\ simulated data samples,  the average bias on 
$w_0$ and $w_a$ is consistent with zero. 
The host photo-$z$ systematic of 0.01 reduces FoM by only 2\% because 
i) most $z<0.5$ events are in the $\zspec$ subset,
ii) the combined SN+host photo-$z$ has $\times 2$ smaller bias, and
iii) the anti-correlation between fitted redshift and color self corrects distance errors.
To prepare for analysing real data, the next SNIa-cosmology analysis with photo-$z$'s 
should include non SN-Ia contamination and host galaxy mis-associations. 
\end{abstract}

\maketitle

\section{Introduction}
\label{sec:intro}
Since the discovery of cosmic acceleration \citep{perl,adam} using Type Ia supernovae (SNe~Ia), this geometric probe has
provided unique constraints on the dark energy equation of state (EOS) today, $w_0$, and its variation with cosmic time, $w_a$ \citep{cpl,cpl2}. The most precise measurements of the dark energy EOS have been based on ${\sim}1000$ \specy\ confirmed SN samples with \spec\ redshifts
from the SN or host-galaxy \citep{Betoule2014,Scolnic2018,DES3YR,pantheon_new}.

Over the next decade, much larger SN samples are expected from  the Vera C. Rubin Observatory and Legacy Survey of Space and Time (LSST\footnote{\URLLSST}) 
and the Nancy Grace Roman Space Telescope.
 Spectroscopic resources will be capable of observing only a small fraction of the discovered SNe.
To make full use of these future samples in cosmology analyses,  well developed methods have been used for 
photometric classification using broadband photometry 
\citep{Lochner2016,Moller2020}. 
 A photometric redshift method  using the SN+host galaxy photo-$z$ has been proposed 
\citep{Kessler2010,Palanque2010,zbeams}, but a rigorous SNIa-cosmology analysis with photo-$z$'s has not been performed. 
   
To analyse SN~Ia samples with contamination from other SN types,
the ``BEAMS''\footnote{BEAMS: Bayesian Estimation Applied to Multiple Species} framework was developed to rigorously use the photometric classification probabilities \citep{kunz,hlozek12}.
The BEAMS framework, combined with photometric classification, 
was first used to obtain SNIa-cosmology results from Pan-Starrs1 data \citep{Jones2018}. 
An extension to BEAMS, 
``BEAMS with Bias Corrections'' (BBC; \citet{bbc}; hereafter KS17), 
was used in \citet{Jones2018} and is currently used in the analysis of data 
from the Dark Energy Survey \citep{Vincenzi2022}.

To analyze SN~Ia samples using photometric redshifts, 
\citet{Kessler2010} and \citet{Palanque2010} extended
the SALT2 light curve fitting framework \citep{salt2} 
to include redshift as an additional fitted parameter,
and to use the host-galaxy photo-$z$ as a prior.
\citet{Dai2018} analyzed a simulated LSST sample including SNe~Ia and SNe CC,
and applied both photometric classification (but not BEAMS) and the SALT2 photo-$z$ method. 
They fit the resulting Hubble diagram with a flat-$\Lambda$CDM model 
and recovered unbiased $\OM$ with a statistical precision of $0.008$.
Using data from the Dark Energy Survey (DES), 
\citet{Chen2022} performed a photo-$z$ analysis using a subset of
${\sim}100$ SNe~Ia hosted by {\tt redMagic} galaxies for which 
both photometric and spectroscopic redshifts are available.
Fitting their Hubble diagram with a flat \wCDM\ model, they find a $w$-difference
of 0.005 between using spectroscopic and photometric ({\tt redMagic}) redshifts.
Finally, \citet{Linder2019, Mitra2021} evaluated the impact of photometric redshifts for LSST
using a Fisher matrix approximation 
that does not include light curve fitting or bias corrections. They concluded that for $z<0.2$, spectroscopic redshifts are necessary for robust cosmology measurements.

A hierarchical Bayesian methodology \citep[zBEAMS]{zbeams} has been proposed to combine 
photometric classification (BEAMS), photometric host-galaxy redshifts, 
and incorrect host-galaxy assignments. This method has been validated on a toy simulation
of SN distances with random fluctuations, 
but the analysis does not include light curve fitting, bias corrections, or systematic uncertainties.

Here we present a rigorous SNIa-cosmology analysis on simulated LSST data that includes  host galaxy photo-$z$'s. We use these photo-$z$s  to include more distant SNe that would otherwise be excluded 
in a \specy\ confirmed sample, and we evaluate the impact of including these additional SNe in the cosmology analysis. 
Our simulation is based on the cadence of the Deep Drilling Fields (DDF) from the 
\textbf{P}hotometric \textbf{L}SST \textbf{As}tronomical \textbf{Ti}me-series 
\textbf{C}lassification \textbf{C}hallenge \citep[\acro,][see sec. \S\ref{plastic}]{plasticc_K2019}, 
combined with a \lowz sample based on the cadence of the Wide Fast Deep (WFD) fields.
Our end-to-end analysis includes light curve fitting, simulated bias corrections applied with BBC, 
a covariance matrix that includes systematic uncertainties, 
and fitting a bias-corrected Hubble diagram for cosmological parameters. We examine the \wCDM\ and \wwCDM\ models.

We adopt the photo-$z$  method from \citet{Kessler2010},
and we use the host-galaxy photo-$z$ as a  prior.
To focus on photo-$z$ issues, we simulate SNe~Ia only
(without contamination) and assume that all host-galaxies are correctly identified.
Therefore, the BEAMS formalism is not used in the analysis.
We use science codes from the publicly available 
\textbf{S}uper\textbf{N}ova \textbf{ANA}lysis software package \SNANA\footnote{\URLSNANA} \citep{snana}, and we use the cosmology-analysis workflow from \pippin\ \citep{pippin}.

This paper is presented as follows. 
In sec.~\ref{sec:lsst}, we 
briefly review LSST and the Dark Energy Science Collaboration.
Sec~\ref{sec:sim} describes the simulated data sample and 
Sec.~\ref{sec:anaysis} describes the cosmology analysis.
Results are presented in sec.~\ref{sec:results} and we conclude in Sec.~\ref{sec:conclude}.

\section{Overview of LSST and Dark Energy Science Collaboration}
\label{sec:lsst}

LSST  
is a ground-based stage IV dark energy survey program \citep{Cahn2009,ivezic}. 
It is expected to become operational in $2023$, 
and will discover millions of supernova over the 10 year survey duration.
The Simonyi Survey optical Telescope at the Rubin Observatory  includes an $8.4~m$ mirror\footnote{$6.7$~m of effective aperture}    and a state-of-the-art $3200$ megapixel camera ($9.6$~deg$^2$ FoV) that will provide
  the deepest and the widest views 
of the Universe with  unprecedented quality. 
LSST  will observe nearly half the night sky every week to a depth of $24^{th}$ magnitude 
in the six filter bands (\bands) spanning wavelengths from ultra-violet to near-infrared.

The Dark Energy Science Collaboration (DESC\footnote{\URLDESC}) 
is an analysis team with nearly 1,000 members, and their goal is to make numerous
high accuracy measurements of fundamental cosmological parameters using data from LSST. 
Prior to first light, DESC has implemented data challenges as a strategy to continuously develop
analysis pipelines. This photo-$z$ analysis within the Time Domain working group
leverages two previous challenges:
1) a transient classification challenge ({\acro}), and
2) an image-processing challenge (DC2: \citet{dc2,Sanchez}).
An updated \acro\ challenge, with several new models and transient-host correlations
\citep{lokken},
is under development to test early classification and to test processing large 
numbers of detection ``alerts'' expected from the Rubin Observatory.

\section{Simulated Data}\label{sec:sim}
We do not work with simulated images and thus we don’t run the LSST difference imaging analysis (DIA)\footnote{\url{https://github.com/LSSTDESC/dia_pipe}} based on \citet{DIA1}. 
Instead, we simulate SN~Ia  light curves corresponding to the output of DIA, and calibrated to the AB magnitude system \citep{fukugita}. 
Following \acro~{\citep{plasticc_K2019,plasticc_H2020}}, we use the cadence and observing properties from MINION1016\footnote{\url{http://ls.st/Collection-4604}} and we
  include a host galaxy photometric redshift and rms uncertainty based on 
\citet[hereafter G18]{Graham2018_photoz}, 
but we do not model correlations between the SNe and host galaxy properties.
\acro\ was designed to motivate the development of  classification algorithms
for photometric light curves from  transients discovered by LSST.  

\acro\ included two LSST observing strategies:
1) five Deep-Drilling-Fields (DDF),
covering ${\sim}50$~deg$^{2}$, that are 
revisited frequently and hence correspond to areas with enhanced depth  and 
2) the Wide-Fast-Deep (WFD) covering a
majority of the southern sky ($18000$ deg$^2$). 
We simulate a high-$z$ sample  using DDF and co-add the nightly observations within each band (sec.~\ref{plastic}).  Since the \acro\ DDF data has limited statistics at low redshifts,
we compliment the \acro\ data with a \specy\ confirmed low-$z$ sample (sec.~\ref{sec:lowz}) based on the wide-fast-deep (WFD) cadence used in DC2.

Rather than using the publicly available \acro\ data, we regenerate the DDF simulation because our analysis needs a much larger sample for bias corrections that is not publicly available. We have verified that our new sample is statistically equivalent to the public data by comparing distributions of redshift, color and stretch. Our simulation does not include contamination from core collapse and peculiar SNe,
nor DIA artifacts such as 
catastrophic flux outliers, PSF model errors, and non-linearites. 

The simulation process adapted in this analysis is described in depth in \citet{kessler2019}. 
To accurately measure biases on cosmological parameters, \NSAMPLE\ statistically independent simulated data samples are 
generated and each sample is analyzed seperately. 

\begin{figure*}[tb]
    \centering
    \makebox[\textwidth]{\includegraphics[width=0.8\paperwidth]{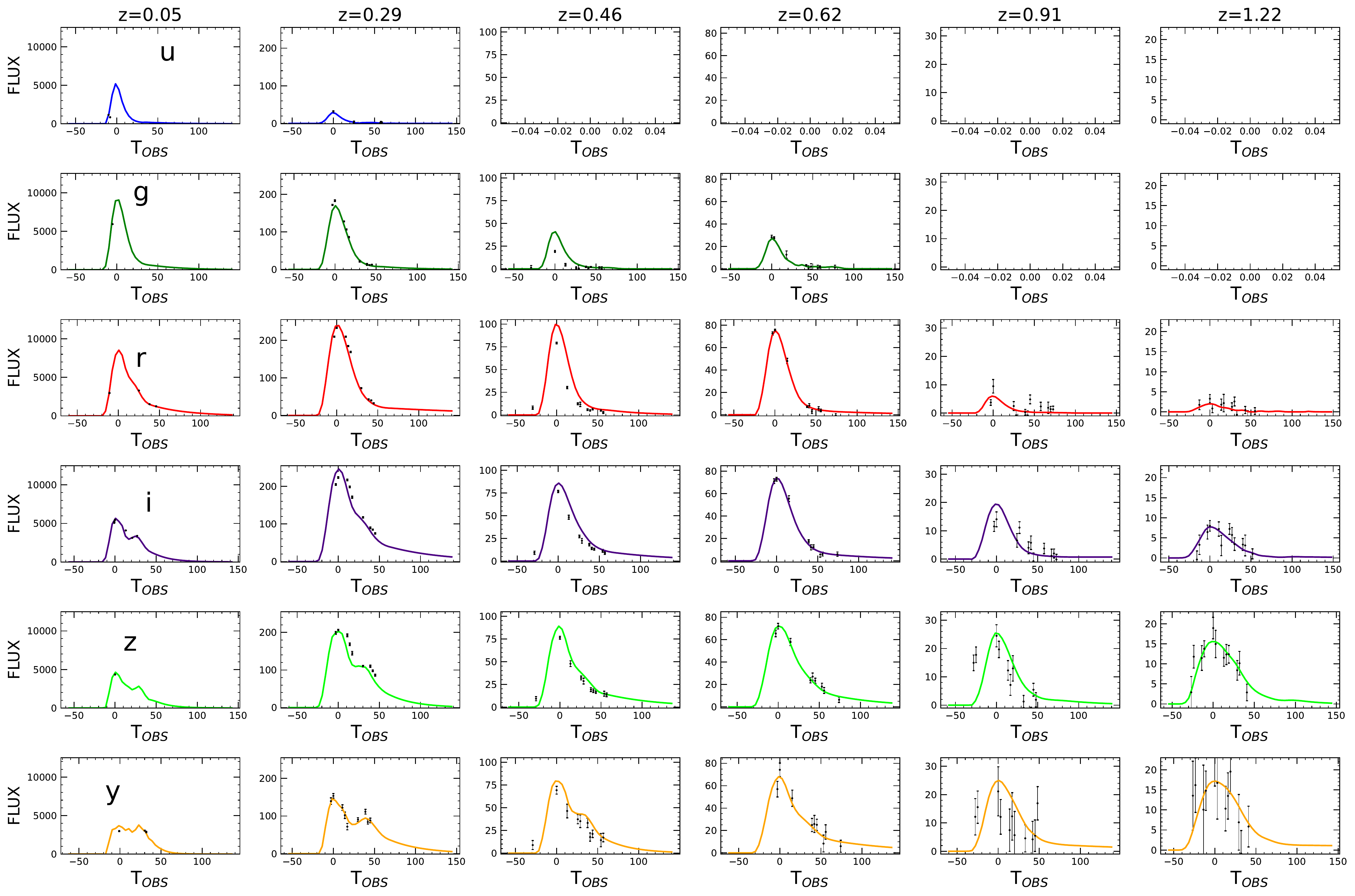}}
    \caption{Sample simulated light curves (calibrated flux vs. 
    $T_{\rm OBS} = \rm MJD - t_0$) 
    for redshifts spanning $z\sim[0-1.2]$. Each column shows a light curve from a single event in each of the  six LSST optical pass band filters $u,g,r,i,z,y$. 
    Left most column ($z=0.05$) is from low-$z$ (WFD); the remaining events are from high-$z$ (DDF). The smooth curves are fits from the SALT2 model, and each color corresponds to a different passband}. 
    \label{fig:lc}
\end{figure*}

A summary of average simulation statistics is shown in Table \ref{T4}.   For the high-z sample, the number of generated events (\Ngen\ column of Table \ref{T4}) is computed from the measured volumetric rate, duration of the survey, and $50$ deg$^2$ area of DDF. For low-$z$, \Ngen\ is  arbitrarily chosen such that the number of 
 events after selection requirements is roughly $500$, 
which is about $\sim 10\%$ of the high-z statistics. 
Examples of simulated light curves at different redshifts 
 are shown by the black circles in Fig.~\ref{fig:lc}.

\begin{table}[ht!]
\begin{center}
\caption{Summary of simulations statistics.}
\begin{tabular}{|c|c|c|c|ccc|}
\hline
&$z$- range & \Ngen\  & \Ngen\ & After&Selection Cuts: & \\
& & total\footnote{Total number of generated SN Ia} & trigger\footnote{Two or more detections separated by more than $30$ minutes}&
        $\zspec$\footnote{Subset of events with spectroscopic redshift. } & 
        full sample &  \\
[0.5ex]
 \hline
     Low-$z$  &$0.01-0.08$ & $4200$  &$696$ & $539$ &  $539$  &  \\
     High-$z$ &$0.03-1.55$ & $41819$ &$12808$ & $1482$& $4873$ & \\
   \hline
\end{tabular}
\label{T4}
\end{center}
\end{table}

\subsection{High-\protect{$z$} data : \acro }
\label{plastic}
The original \acro\ simulation covers the first three years of LSST
with 18 models that include both extragalactic and galactic transients.
For this analysis, we simulate only SNe~Ia using the SALT2 model \citep{salt2}.
This model includes 
measured populations of stretch and color from \citet{SK2016} with
stretch- and color-luminosity parameters 
($\alpha=\alphaTrueVal$, $\beta=\betaTrueVal)$,
an intrinsic scatter of the model SED, and 
a near-infrared extension \citep{salt+} 
to include  the  $i,z,y$ band wavelength range.
Correlations between SNe and host-galaxy mass are not included.
Next, the model SED is modified to account for
cosmic expansion ($\OM=0.315$, $w=-1$, flatness) redshift, and Galactic extinction from  \citet{ebv2}. 
Filter passbands are used to compute broadband fluxes at epochs determined
by the DDF cadence from  \verb|OpSim|  \citep{OpSim1,OpSim2,OpSim3}, 
and observing conditions (zero-point, PSF and sky noise) are used to model flux uncertainties. 
The $5\sigma$ limiting magnitudes for each of the \bands\ passbands are listed in Table~\ref{t_lowz} 
for both the low-$z$ and the high-$z$ samples. 
We adopt the detection efficiency vs. signal-to-noise ratio (SNR) from the DC2 analysis as shown
in Fig.~9 of \citet{Sanchez}.
The simulated trigger selects events with two detections separated by at least 30 minutes.

\begin{figure}[h]
    \centering
    \subfloat{{\includegraphics[width=.394\textwidth]{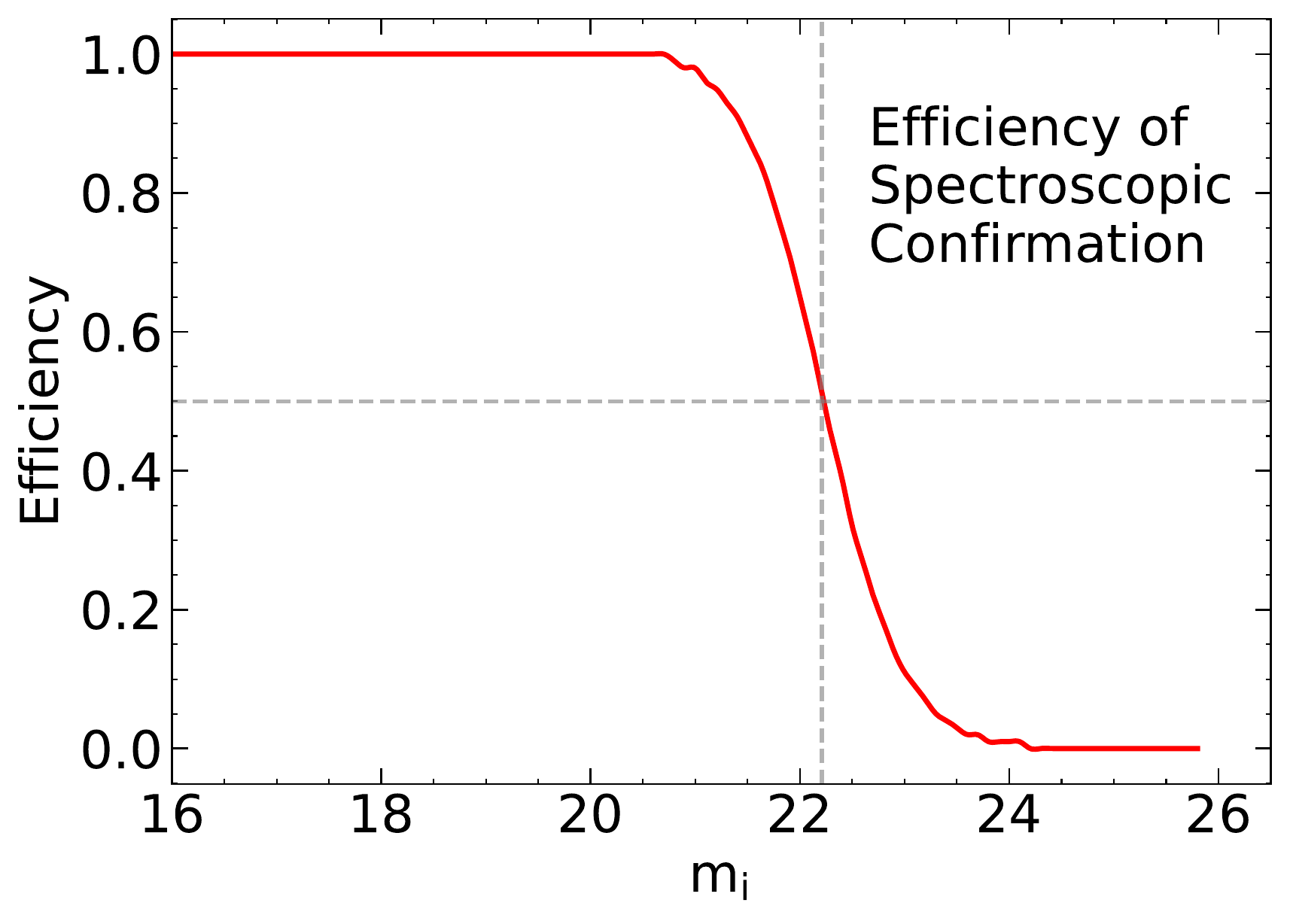} }}

    \subfloat{{\includegraphics[width=.408\textwidth]{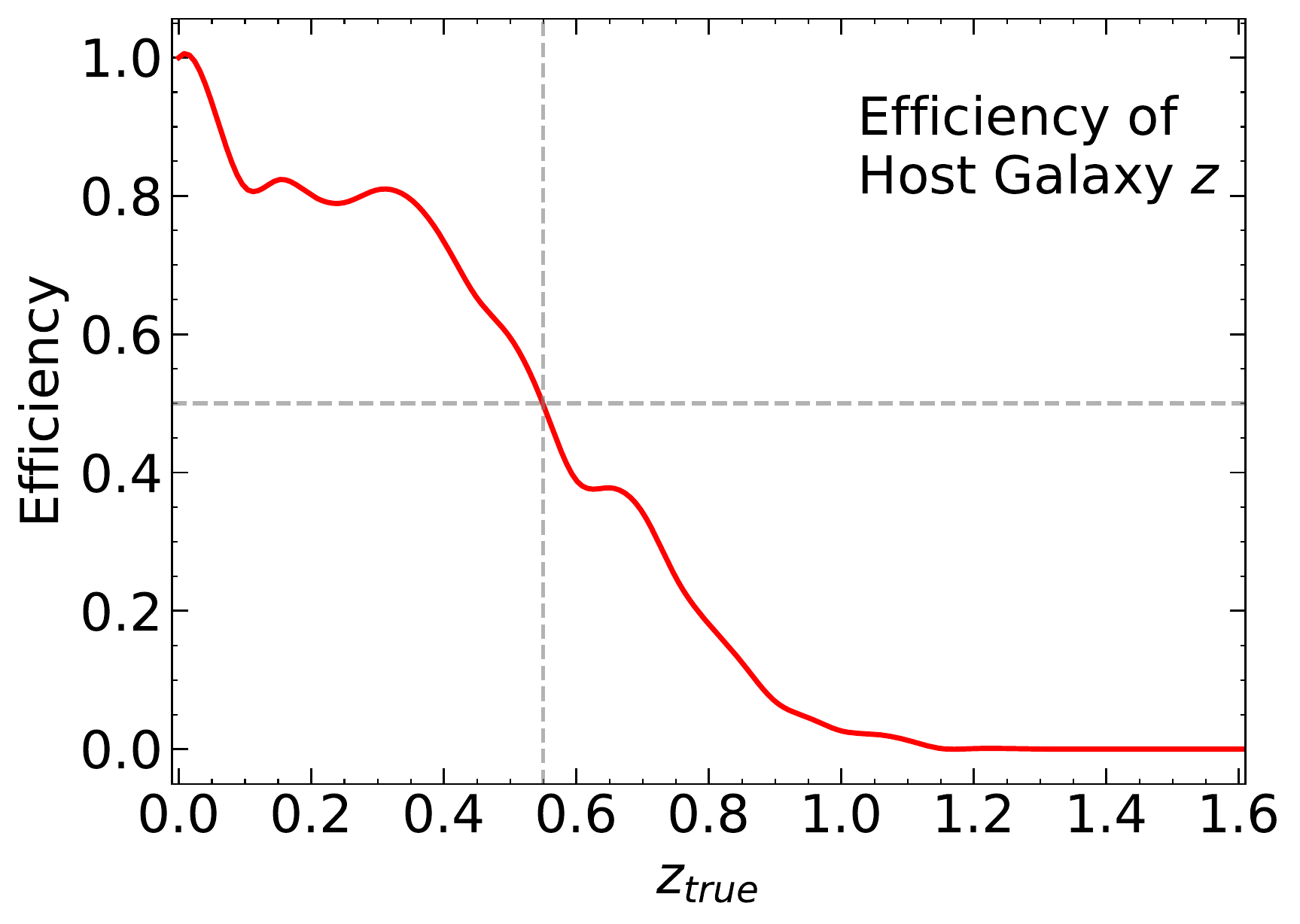} }}
     \caption{{For DDF, Top : Efficiency vs. peak $i$-band magnitude ($m_i$)  for the spectroscopically confirmed events. Bottom : Simulated efficiency vs. redshift for measuring a spectroscopic host galaxy redshift.
     The grey dashed lines show the $50\%$ efficiency: $m_i=22.2$ (top), and $\ztrue=0.56$ (bottom).}
     }
    \label{fig:zeff}
\end{figure}

Following \acro, we define a ``$\zspec$'' sample consisting of  two  subsets of events with accurate spectroscopic redshifts ($\sigma_{z}
\sim 10^{-5}$). The first subset assumes an accurate redshift from  
 \specy\ confirmed events based on a forecast of the performance for the 4-metre Multi-Object Spectroscopic Telescope spectrograph
\citep[4MOST hereafter, ][]{4MOST2}\footnote{\url{https://www.4most.eu/cms}} that is under construction by the European Southern Observatory (ESO)\footnote{https://www.eso.org/public/}. 
4MOST is expected to begin operation in $2023$ (similar to the LSST timeline) and is
located at a latitude similar to that of the Rubin observatory in Chile.  
The second subset includes photometrically identified events with an accurate 
host galaxy redshift using 4MOST. 
The second subset has about $\sim60$\% more events than the first subset, 
and each subset is treated identically in the analysis. 
The simulated efficiency vs. redshift  for each $\zspec$ subset is shown in Fig.~\ref{fig:zeff}.

\begin{figure*}[!ht]
  \begin{minipage}{\textwidth}
    \centering
    \includegraphics[width=.4\textwidth]{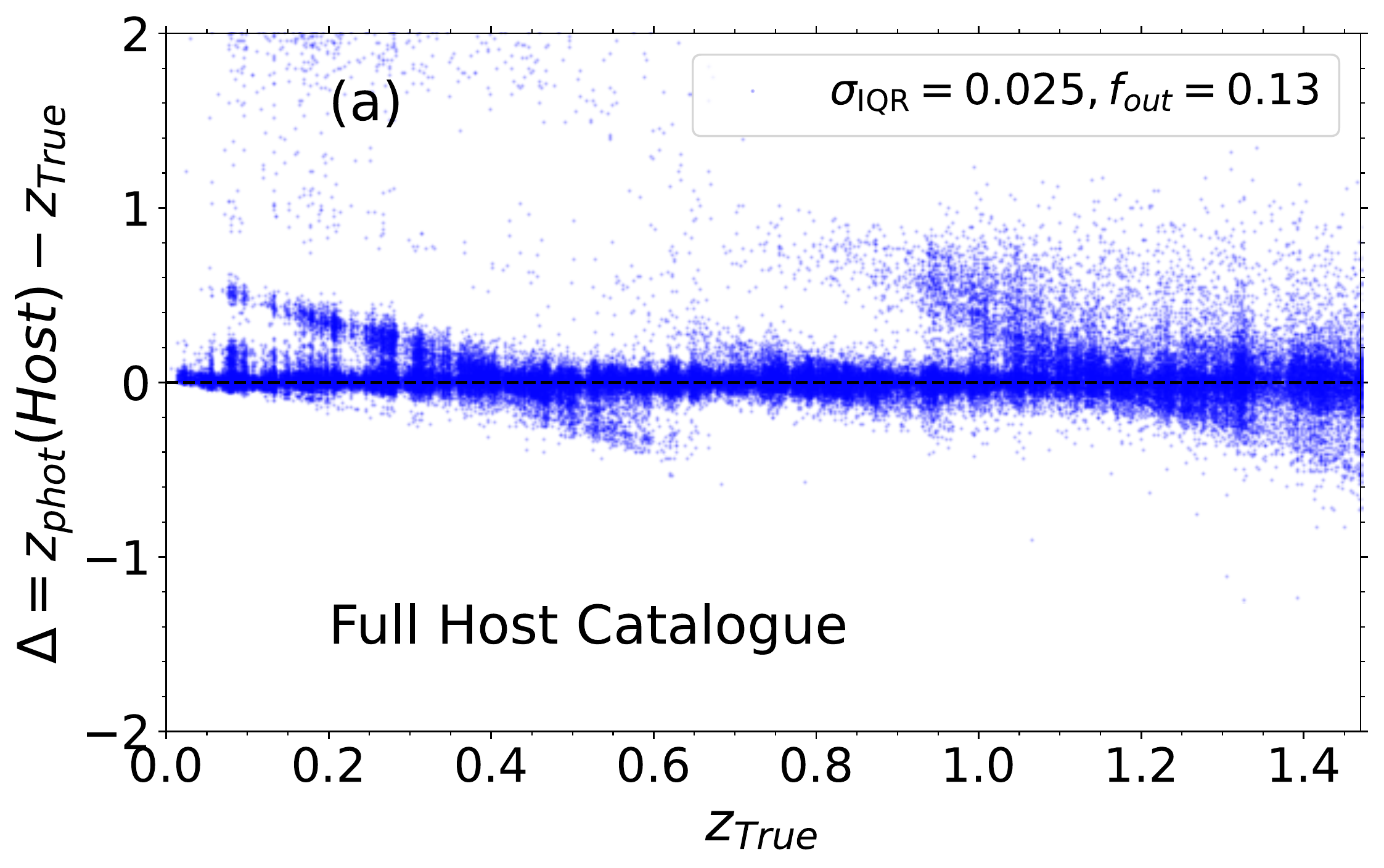}\quad
    \includegraphics[width=.4\textwidth]{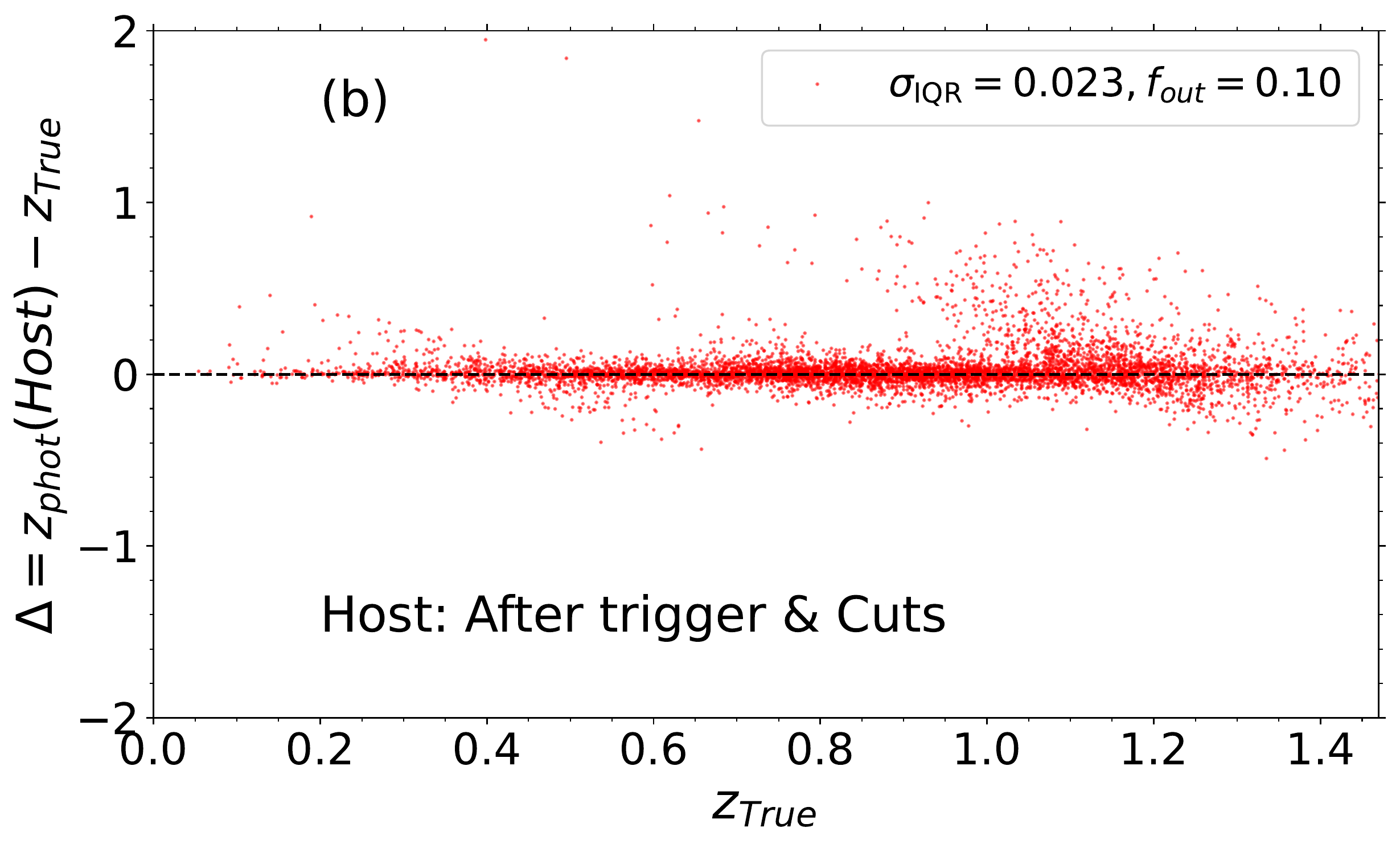}\\
    \includegraphics[width=.4\textwidth]{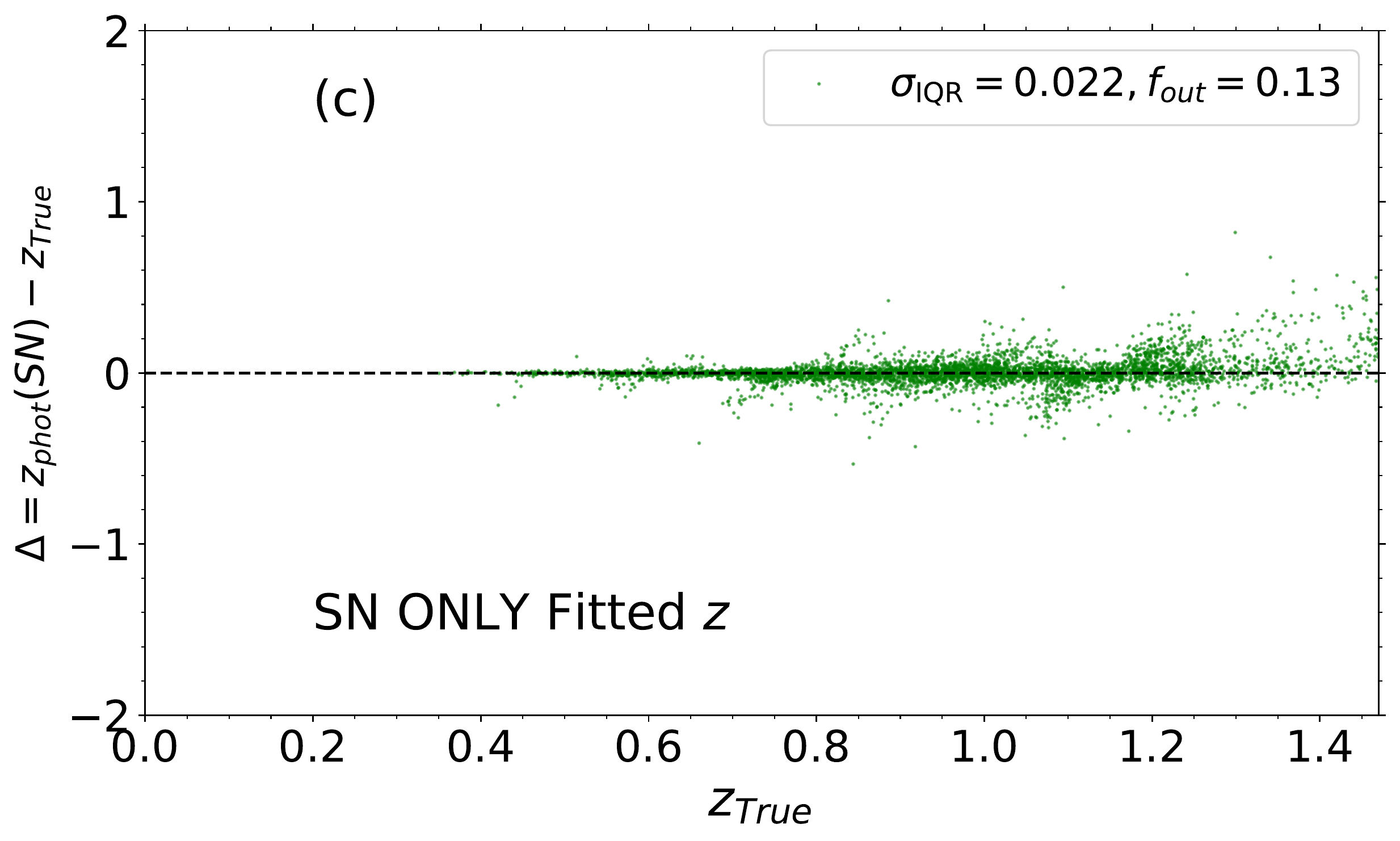}\quad
    \includegraphics[width=.4\textwidth]{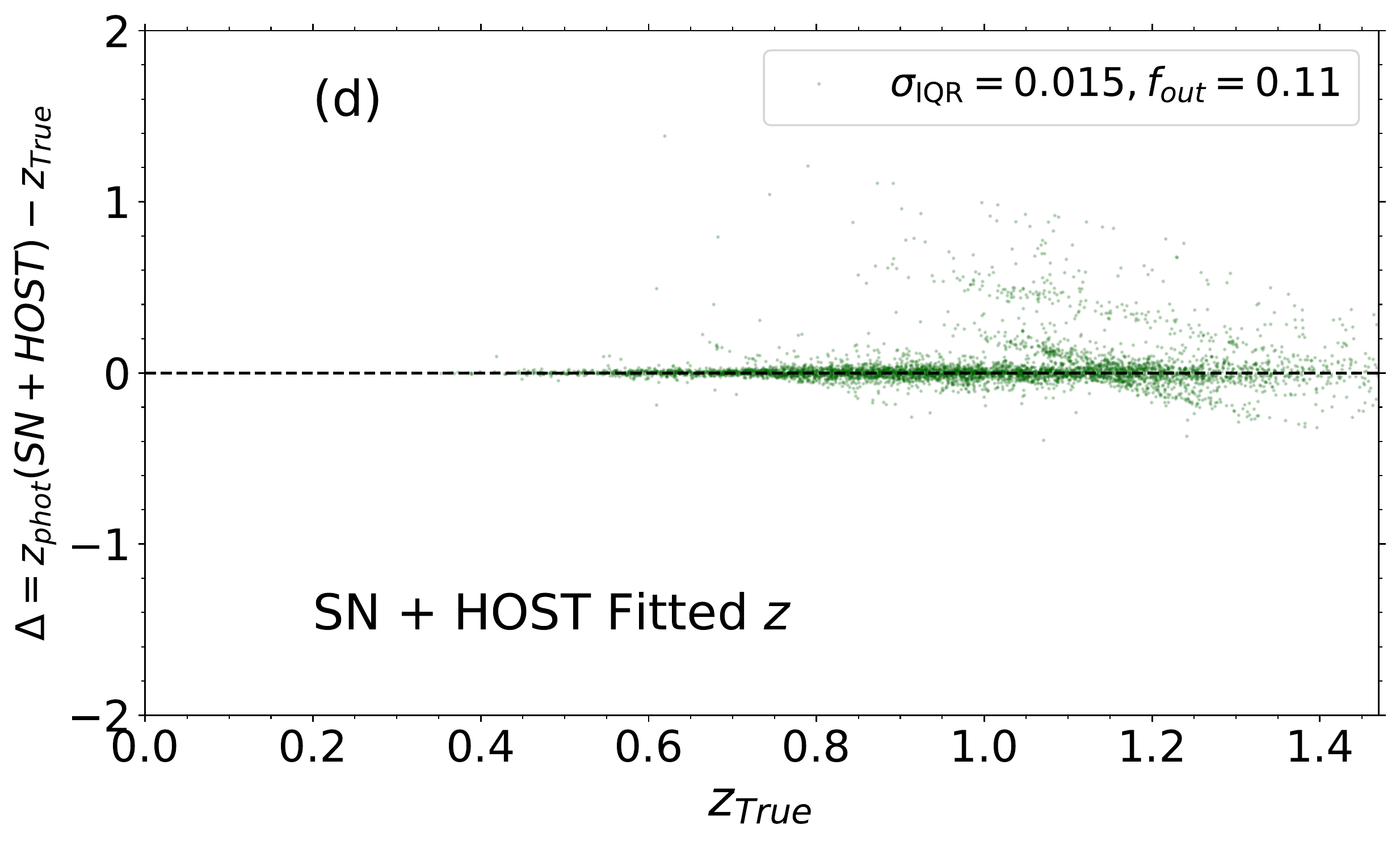}
    \caption{Photo-$z$ residual ($\zphot-\ztrue$) vs. $\ztrue$ for
  (a) full host galaxy catalogue, 
  (b) host galaxy after trigger and selection cuts,  
  (c)
   SALT2 fitted SN only photo-$z$  without host galaxy prior and 
   (d) Combined SALT2 fitted SN+Host photo-$z$ that  is used for the Hubble diagram. 
   Panels (b), (c) and (d) have no $\zspec$ events. 
   The $\sigIQR$ and $\fout$ numbers on each panel are computed for $0.4<\ztrue < 1.4$.}
    \label{fig:photoz_res}
  \end{minipage}\\[1em]
\end{figure*}

To estimate the host-galaxy photometric redshifts, \acro\ used a 
\textbf{C}olor \textbf{M}atched \textbf{N}earest \textbf{N}eighbour photometric redshift estimator
(CMNN in G18). 
CMNN uses a five-dimensional color space grid to train  a set of galaxies and defines a distance metric that is used on the test set to assign the redshift and the associated uncertainty. 
Figure \ref{fig:photoz_res}{a} shows  the 
photo-$z$ residuals, $\zphot-\ztrue$, as a function of $\ztrue$.

To characterise the residuals, we follow \citet{Graham2018_photoz} and define metrics for 
an inner core resolution and outlier fraction using the quantity $\Dz = |\zphot-\ztrue|/(1+\zphot)$.
The resolution is the width of the inter quantile distribution of $\Dz$, divided by 1.349,
and is denoted by $\sigIQR$. The outlier fraction ($\fout$) is the fraction of events satisfying

%
\begin{equation}
     \Dz > 3\sigIQR \ \rm{and}\  \Dz > 0.06 ~~.
      \label{eq:out1}
\end{equation}
For $\ztrue<0.4$ the events have a \spec\ redshift, and for $\ztrue>1.4$ the SNe are too faint for detection.
For the relevant redshift range ($0.4 < \ztrue < 1.4$), 
$\sigIQR=0.025$ and $\fout=0.13$.

Following \acro, we use the volumetric rate  model $R(z)$ 
based on \citet{Dilday2008} for $z<1$ and \citet{Hounsell2018} for $z>1$. The rate $R(z)$ we adopt is given by
\begin{eqnarray}
     R(z)  & = & 2.5  \times 10^{-5} (1+z)^{1.5}~{\RateUnit} ~~(z<1) \\
     R(z) & = & 9.7  \times 10^{-5} (1+z)^{-0.5}~{\RateUnit} ~~(z>1)~.
\end{eqnarray}

\subsection{Low-z data : Spectroscopic }
\label{sec:lowz}
We simulate a \specy\ confirmed low-$z$ sample based on the WFD cadence from DC2.
We assume accurate \spec\ redshifts
and  100\% efficiency up to redshift $z<0.08$. 
The simulation code and SNIa model are the same as for the high-$z$ sample. 
Compared to DDF, the WFD cadence has $30\%$ fewer observations on average
and has $1$~mag shallower depth (Table~\ref{t_lowz}).

\begin{table}
\begin{center}
\caption{Average depth and time between observations.}
\begin{tabular}{ | c | c  c | c  c |} 
 \hline
          & \multicolumn{2}{c|}{WFD} & \multicolumn{2}{c|}{DDF} \\
 Filter   & depth\tablenote{$5\sigma$ limiting magnitude.}   
          & gap\tablenote{Average time (days) between visits, excluding seasonal gaps.}         & depth   & gap \\   
 \hline
    $u$ & $23.84$ & $10.5$  &  $25.05$ & $5.3$\\
 \hline
   $g$  & $24.80$ & $11.9$&  $25.52$ & $7.3$\\
 \hline
    $r$ &$24.21$ &$8.2$ & $25.60$ &$7.3$ \\
 \hline
   $i$ & $23.57$ & $8.6$&  $25.19$ & $7.3$\\
 \hline
    $z$&$22.65$  &$9.0$      &$24.79$  &$7.3$ \\
 \hline
    $Y$&$21.79$  &$11.2$  &$23.83$  &$7.4$ \\
 \hline
\end{tabular}
\label{t_lowz}
\end{center}
\end{table}

\section{Analysis}
\label{sec:anaysis}

\begin{figure*}[tb]
    \centering
    \makebox[1.\textwidth]{\includegraphics[width=1.0\textwidth]{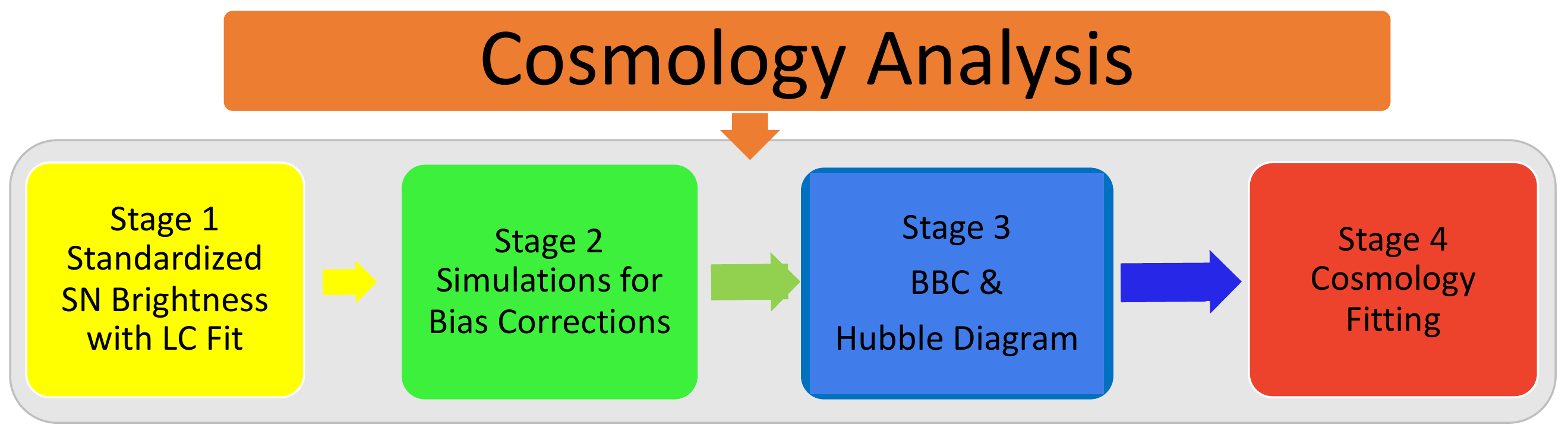}}
    \caption{ Flowchart showing the cosmology analysis steps. }
    \label{fig:analysis}
\end{figure*}

The SNIa-cosmology analysis steps are shown in Fig.~\ref{fig:analysis},
and described below. This analysis is similar to the recent DC2-SNIa cosmology
analysis in \citet{Sanchez}, except here we  include DDF and use photo-$z$ information.
The analysis is performed three times, each using the same low-$z$ sample but varying the high-$z$ data: 
\vspace{-0.2cm}
\begin{enumerate}
 \setlength\itemsep{0.0em}
    \item $\zphot$: full sample including both spectroscopic and photometric redshifts
    \item $\zspec$: subset with only accurate \spec\ redshift from either the host galaxy or SN
    \item $\zcheat$: full sample forcing $\zphot=\ztrue$
\end{enumerate}

\subsection{Lightcurve Fitting and Selection Requirements}
\label{subsec:ana_lcfit}

To standardize the SNIa brightness, we fit each light curve to the SALT2 light curve model 
\citep{Guy2010}, which determines the 
time of peak brightness ($t_0$), amplitude ($x_0$), stretch ($x_1$), and color ($c$).
Previous cosmology analyses have all used SNe with accurate $\zspec$,
and thus redshift had always been a fixed parameter in the SALT2 fit.
In our analysis, the SALT2 fit uses the methodology in \citet{Kessler2010} in which 
the redshift is floated as a  5th parameter, which we call  ``$\zphot$''. 
The host-galaxy photo-$z$ is used as a prior in the SALT2 fit, approximated by a Gaussian 
with mean and $\sigma$
corresponding to the mean and rms of the photo-$z$ PDF.
For the subset with accurate $\zspec$,
the redshift prior is so precise ($0.0001$) 
that such fits are essentially equivalent to fixing the redshift in a 4-parameter fit. Note that $\zphot$ refers to the fitted redshift for all events, including the $\zspec$ subset.

We apply the following selection requirements (cuts) based on analyses using real data:
\begin{enumerate}
  \setlength\itemsep{0.1em}
   \item at least three bands with maximum SNR$>4$ 
   \item successful light curve fit
   \item $|x_1| < 3.0$
   \item $|c| < 0.3$
   \item stretch uncertainty $\sigma_{x1} < 1.0$
   \item time of peak brightness uncertainty $\sigma_{t0} < 2.0$~days
   \item $\Pfit > 0.05$\footnote{$\Pfit$ is the SALT2 fit probability 
          computed from $\chi^2$ and number of degrees of freedom.}
   \item $0.01 < \zphot < 1.4$
   \item valid bias correction  (see  section \S\ref{subsec:ana_bbc}).
\end{enumerate}

SALT2 light curve fits for several events are shown by the smooth curves in Fig.~\ref{fig:lc}.
After selection requirements, the redshift distribution is shown in Fig.~\ref{fig:musig_compare}{a}
 for the subset with and without $\zspec$.

The  $\zphot$ residual vs. $\ztrue$ is shown in Fig.~\ref{fig:photoz_res}\textcolor{red}{a}
for all galaxies in the catalogue, 
and in Fig.~\ref{fig:photoz_res}\textcolor{red}{b} for host galaxies
after SN~Ia trigger and selection cuts.
After selection cuts, 
SNe associated with host-galaxy photo-$z$  outliers tend to be excluded by the SALT2 fit and $\Pfit$ cut;
the core resolution is reduced
by $10\%$, and the outlier fraction is reduced by 20\%.

To compare the photo-$z$ precision between the host and SN, we performed SALT2 light curve fits without
a host-galaxy photo-$z$ prior to determine the SN-only $\zphot$ residuals
(Fig.~\ref{fig:photoz_res}c); the SN-only $\zphot$ core resolution is slightly $({\sim}1.1)$ better than
for the galaxies in Fig.~\ref{fig:photoz_res}a, although the outlier fractions are the same.
For the combined SN+host SALT2 fits, Fig.~\ref{fig:photoz_res}d shows 
$\zphot$  residuals vs. $\ztrue$;
compared to  fitting SN-only,
the SN+host $\zphot$ resolution is 30\% smaller
and has $\sim15\%$ fewer outliers.

To evaluate systematic uncertainties, the SALT2 light curve fits and BBC fit are repeated \NSYST\ times, each   with a separate 
variation shown in Table~\ref{tb:syst_sources}. Each variation results in a distance modulus variation, and we compute a systematic covariance matrix  ($\COVsyst$) using Eq.~6 in \citep{conley11}.

We include variations in 
Galactic extinction, 
calibration,
$\zspec$, and host-galaxy $\zphot$.
We do not include SALT2 modelling and training uncertainties, 
nor do we include uncertainties on the stretch and color populations.

The galactic extinction uncertainty (Row $2$ in Table \ref{tb:syst_sources})  is $\sigma_{\rm{MWEBV}}=0.05\cdot  \rm{MWEBV}$, and is taken from the Pantheon analysis \citep{Scolnic2018}.
The HST calibration uncertainty (Row $3$) is from the  DES SNIa-cosmology analysis 
(Table 4 in \citet{Brout2019_DES3YR_ANA}) and is based on \citet{bohlin:calibrationHST}. 
The zero point uncertainty (Row 4) is  from the LSST science roadmap 
(section \S 3.3 in \citet{LSST_scibook}),
and is consistent with the Pan-STARRS $3\pi$ internal calibration 
accuracy \citep{Schlafly2012, PS1cal2013}.
The wavelength calibration uncertainty (Row 5) is from the Pantheon analysis
in \citet{Scolnic2018}.

The spectroscopic redshift uncertainty (Row 6) is from Table~4 in \citet{Brout2019_DES3YR_ANA},
which is based on low-redshift constraints on local density fluctuations \citep{Calcino2017}.
For the host-galaxy photo-$z$ bias uncertainty (Row 7),
the statistical bias in our \acro\ simulation is well below 0.01 as shown in 
the lower panel of Fig.~2 in G18. 
This statistical bias is valid for the galaxy training set, 
but the bias for the subset of SN~Ia host galaxies is likely to be larger.
Without a photo-$z$ bias estimate for SN~Ia host-galaxies,
we make an ad-hoc estimate from the DES weak lensing (WL) cosmology analysis 
in which \citet{PZ2021_DES3YR_WL} find a statistical $\zphot$ bias of ${\sim}0.001$,
while their {\it weighted} $\zphot$  bias is 0.01, an order of magnitude larger. 
We use their weighted $\zphot$ bias of 0.01 as the systematic uncertainty.
The uncertainty in the host $\zphot$ uncertainty (Row 8) is from the variation in robust standard deviations
in the upper panel of Fig~2 in G18.

\begin{table}
\caption{Source of Systematic Uncertainty } \vspace{-0.5cm}
\begin{center}
\begin{tabular}{ | l | l | l | c | } 
 \hline
Row  & Label & Decription & Value \footnote{Shift (or scale) applied to simulated data before each re-analysis} \\
 \hline\
 1 & StatOnly & no systematic shifts &  --- \\
 \hline\
 2 &  MWEBV & shift $E(B-V)$ & 5\%  \\ 
 \hline\
 3 &  CAL\_HST  & HST  calibration offset & $0.007\times\lambda$\\ \
  4 & CAL\_ZP & LSST zero point shift & $5$ mmag \\ \
5 & CAL\_WAVE & LSST Filter shift & $5$ \AA\\
  \hline\
6 & zSPEC  & shift $\zspec$ redshifts  &   $5\times 10^{-5}$\\ \
 7 & zPHOT  & shift $\zphot$ redshifts  &   $0.01$ \\ \
 8 & zPHOTERR & scale host $\zphot$ uncertainty &  $1.2$\\ 
 \hline
\end{tabular}
\label{tb:syst_sources}
\end{center}
\end{table}

\subsection{Simulated Bias Corrections}
\label{subsec:ana_simbias}
To implement distance bias corrections in BBC (\S\ref{subsec:ana_bbc}), 
we generate a large sample of 
\NBIASCORTOT\ events
 (after cuts, in section \ref{plastic}) 
 which consists of  \NBIASCORHIZ\ high-$z$  events and \NBIASCORLOZ\ low-$z$  events. The bias correction is applied independently for high-$z$ and low-$z$, and thus the relative number of events in each sub-sample need not match the data. 
The simulation procedure is identical to that used for the simulated data, 
except for $\alpha$ and $\beta$. While fixed values are used for the data sample, 
a $2\times2$ $\alpha,\beta$ grid is used for the 
``biasCor'' simulation to enable interpolation in BBC.

\subsection{BEAMS with Bias Corrections (BBC)}
\label{subsec:ana_bbc}
BBC reads the SALT2 fitted parameters (high-$z$ and low-$z$) 
from the data and biasCor simulation,
and produces a bias-corrected Hubble diagram, both unbinned and in redshift bins.
For each event, the measured distance modulus is based on \citet{Trip1998},
\begin{equation}
   \mu = m_B + \alpha x_1 - \beta c  + M_0 + \mubias~,   \label{eq:mu}
\end{equation}
where $m_B \equiv -2.5\log_{10}(x_0)$, 
$\{\alpha,\beta,M_0\}$ are global nuisance parameters, and
$\mubias = \mu - \mutrue$ is determined from the biasCor simulation
in a 5-dimensional space of $\{z,x_1,c,\alpha,\beta\}$. 
A valid bias correction is required for each event, resulting in a few percent loss.
The distance uncertainty ($\sigmu$) is computed from Eq.~3 of KS17.
Since there is no contamination from non-SNIa, all SN~Ia classification probabilities are
set to 1 and we do not use the BEAMS formalism.

There are two subtle issues concerning the use of $\zphot$ and its uncertainty $\sigz$.
First, 
the calculated distance error from $\sigma_z$ ($\sigmu^z$ in Eq.~3 of KS17) 
is an overestimate because it does not account for the correlated color error that 
reduces the distance error.
By floating $\zphot$ in the SALT2 fit, redshift correlations propagate to the other SALT2 parameter uncertainties,
and therefore we set $\sigmu^z=0$.
The second issue concerns the $\mu_\mathrm{{bias}}$ computation, where
$\mutrue$ is computed at SALT2-fitted $\zphot$ rather than the true redshift.

To avoid a dependence on cosmological parameters, the BBC fit is performed in \NZBIN\ logarithmically-spaced redshift bins.
The fitted parameters include the global nuisance parameters 
($\alpha,\beta,M_0$) 
and bias-corrected distances
in \NZBIN\ redshift bins. The unbinned Hubble diagram is obtained from Eq.~\ref{eq:mu} using
the fitted parameters.

 If the same selection requirements are applied to each systematic variation for computing $\COVsyst$,
small fluctuations in the fitted SALT2 parameters and redshift result in slightly different samples,
and these differences introduce statistical noise in  $\COVsyst$.
We avoid this covariance noise by defining a baseline sample for
events passing cuts without systematic variations, and use this same 
baseline sample for all systematic variations. For example, if an event has fitted SALT2
color parameter $c=0.299$, and migrates to $c=0.3001$ for a calibration systematic,
this event is preserved  without applying cuts that require $|c|<0.3$.

To avoid sample differences from the valid bias-correction requirement,
the BBC fit is run twice in which the second fit only includes events 
that have a valid bias correction in all systematic variations.
Finally, for redshift systematics that result in migration to another redshift bin, the original (no syst) redshift bin is preserved for the BBC fit.

\subsection{Cosmology Fitting and Figure of Merit}
\label{subsec:ana_wfit}
For cosmology fitting, we use a fast minimization program that approximates a
CMB prior using the $R$-shift parameter (e.g., see Eq. 69
in \citet{komatsu}) computed from the same cosmological 
parameters that were used to generated the SNe~Ia. The $R$-uncertainty is
$\sigR=0.006$, tuned to have the same constraining power as \citet{Planck2018}.
We fit with  \wCDM\ and \wwCDM\ models, where $w=[w_0+w_a(1-a)]$.
The statistical+systematics covariance matrix is used.
We fit both binned and unbinned Hubble diagrams.

For the \wwCDM\ model, the FoM is computed based on the dark energy task force 
(DETF) definition in \citet{FoM},
\begin{equation}
   \text{FoM} \simeq \frac{1}{ \sigma (w_a) \sigma(w_0) \sqrt{1-\rho^2} }~,
    \label{eq:FoM}
\end{equation} 
where $\rho$ is the reduced covariance between $w_0$ and $w_a$.

\section{Results}\label{sec:results}

For one of the \NSAMPLE\ statistically independent samples, we show the $\zspec$ and $\zphot$ Hubble diagram produced by the BBC fit,
both binned and unbinned, in Fig.~\ref{fig:HD}. The Hubble residuals with respect to the
true cosmology, $\Delta\mu = \mu - \mu_{\rm true}$, are consistent with zero and do not show
a redshift-dependent slope.

\begin{figure}[h]
    \centering
    \includegraphics[width=\columnwidth]{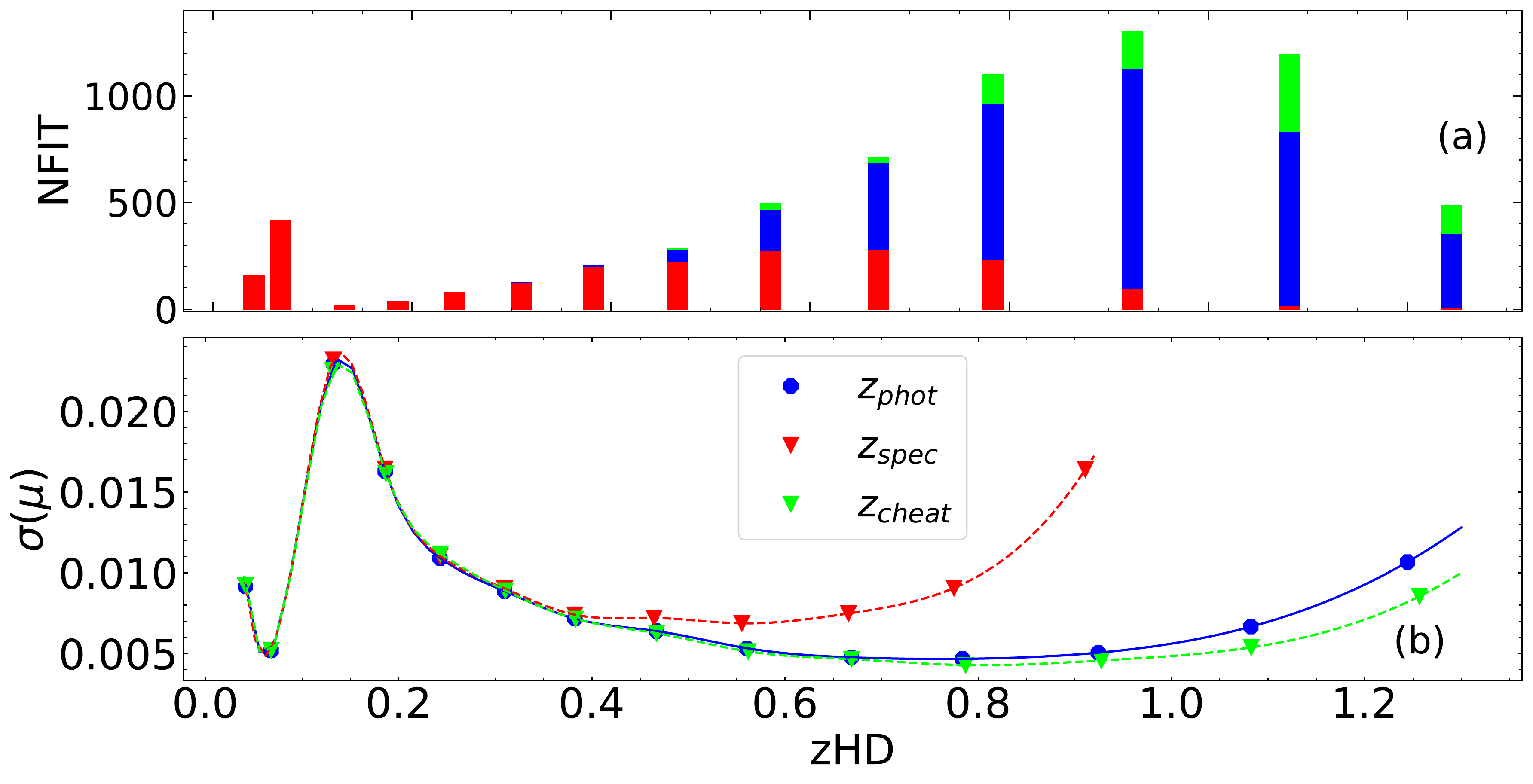}
    \caption{
    Number of events (top) and BBC-fitted distance uncertainty (bottom) per redshift bin. 
     The three sets of overlaid plots correspond to $\zspec$ (red), $\zphot$ (blue), and $\zcheat$ (green). 
    }
    \label{fig:musig_compare}
\end{figure}

The BBC-fitted nuisance parameters are shown in Table~\ref{tb:alphaBeta}
for the three analyses: $\zspec$, $\zphot$, $\zcheat$.
Averaging over the \NSAMPLE\ samples, $\alpha$ and $\beta$ agree well with the simulated inputs.
There is no true $\sigint$ for comparison, but we note that the $\sigint$ values agree
well among the three analyses.

\begin{table}[h]
\begin{center}
\caption{Bias for BBC Fitted Nuisance Parameters\footnote{Averaged over \NSAMPLE\ samples.}}
\begin{tabular}{ | c | c  c | c  c | c |} 
 \hline
 Sample   & $\alpha-\alphaTrueSym$\footnote{$\alphaTrueSym=\alphaTrueVal$}
          &       & $\beta-\betaTrueSym$\footnote{$\betaTrueSym=\betaTrueVal$} 
          &   & $\sigint$ \\    
 \hline
    $\zspec$ & $0.00015\pm 0.00158 $ &   &  $-0.00011\pm 0.02035$ && $0.095$ \\
 \hline
   $\zphot$  & $0.00033\pm 0.00121$ & & $-0.00017\pm 0.01624$ && $0.096$  \\
 \hline
    $\zcheat$ &$0.00049\pm 0.00112$ & & \ \ $0.00125\pm 0.01549$ && $0.095$ \\
 \hline
\end{tabular}
\label{tb:alphaBeta}
\end{center}
\end{table}

Next, we compare the BBC fitted distance uncertainties ($\sigmu$) in redshift bins 
(Fig.~\ref{fig:musig_compare}{b})
The $\zspec$ and $\zphot$ uncertainties are similar for $z<0.5$, 
and at higher redshifts the $\zphot$ uncertainty is significantly smaller than for $\zspec$.
In addition to smaller distance uncertainties, the $\zphot$ redshift range
extends ${\sim}0.3$ beyond that of the $\zspec$ range.

At high redshift, the $\zcheat$ analysis shows little improvement over the $\zphot$ analysis.
Defining an effective distance uncertainty per event in each redshift bin as 
$\sigmubar_z = {\sigmu}_z \times \sqrt{N_z}$, where $N_z$ is the number of events in the redshift bin,
the $\sigmubar_z$ values for $\zcheat$ and $\zphot$ are the same to within a few percent.
There are fewer $\zphot$ events (compared to $\zcheat$) because of selection cuts and
unstable results between multiple light curve fit iterations.


For the cosmology fitting, we fit the binned distances from the BBC fit and also performed
unbinned fits to reduce the systematic uncertainty as described in \citet{Binning_is_sinning}.
While the unbinned cosmology fits result in smaller uncertainties, we find a significant bias
that is driven by the calibration  systematics. 
We have not found an explanation of this bias,
and therefore we present results only for binned distances.

\begin{figure}
    \centering
    \subfloat{%
    \includegraphics[width=0.9\columnwidth]{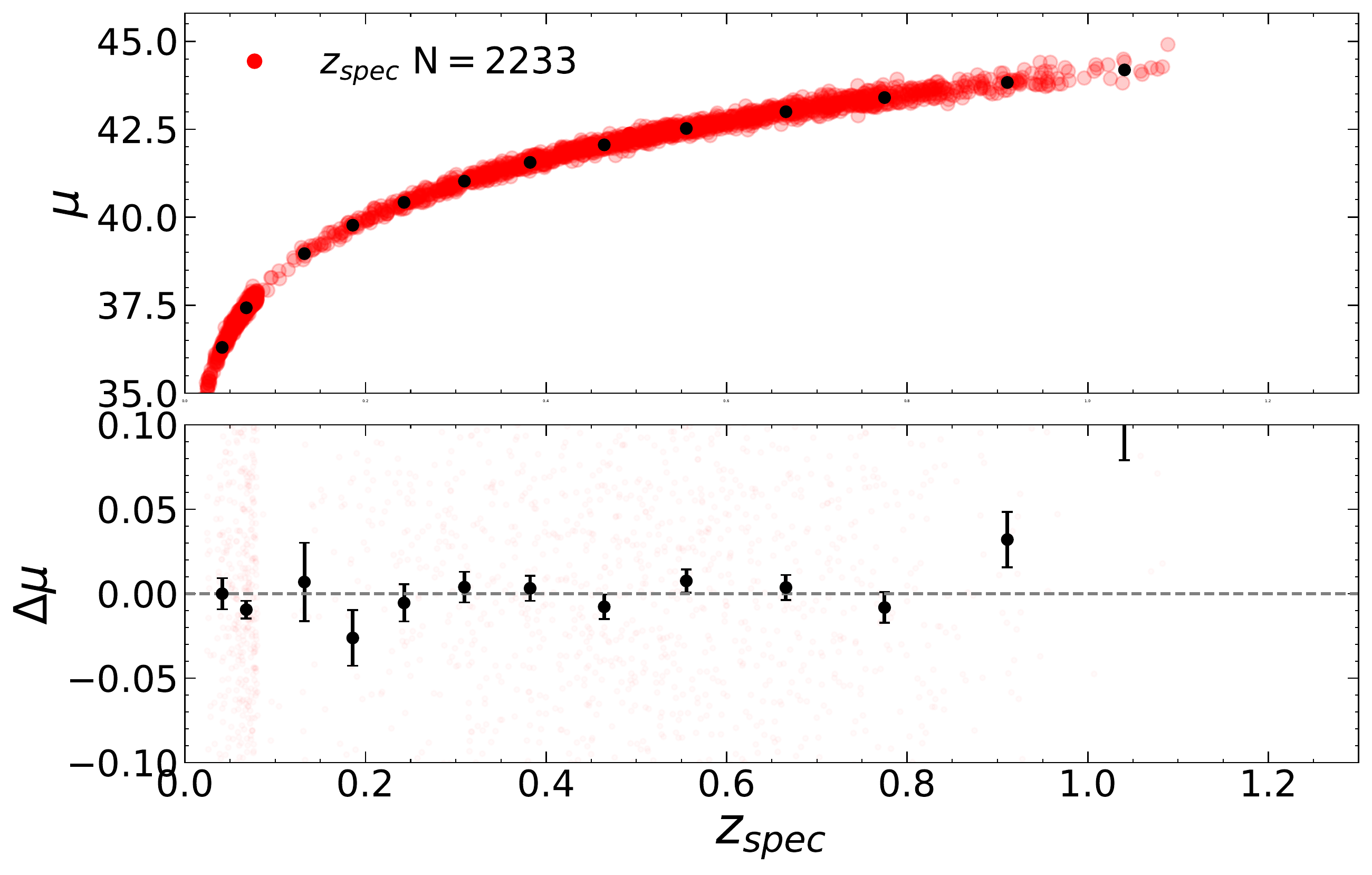}%
    \label{}%
    }
    
    \subfloat{
    \includegraphics[width=0.9\columnwidth]{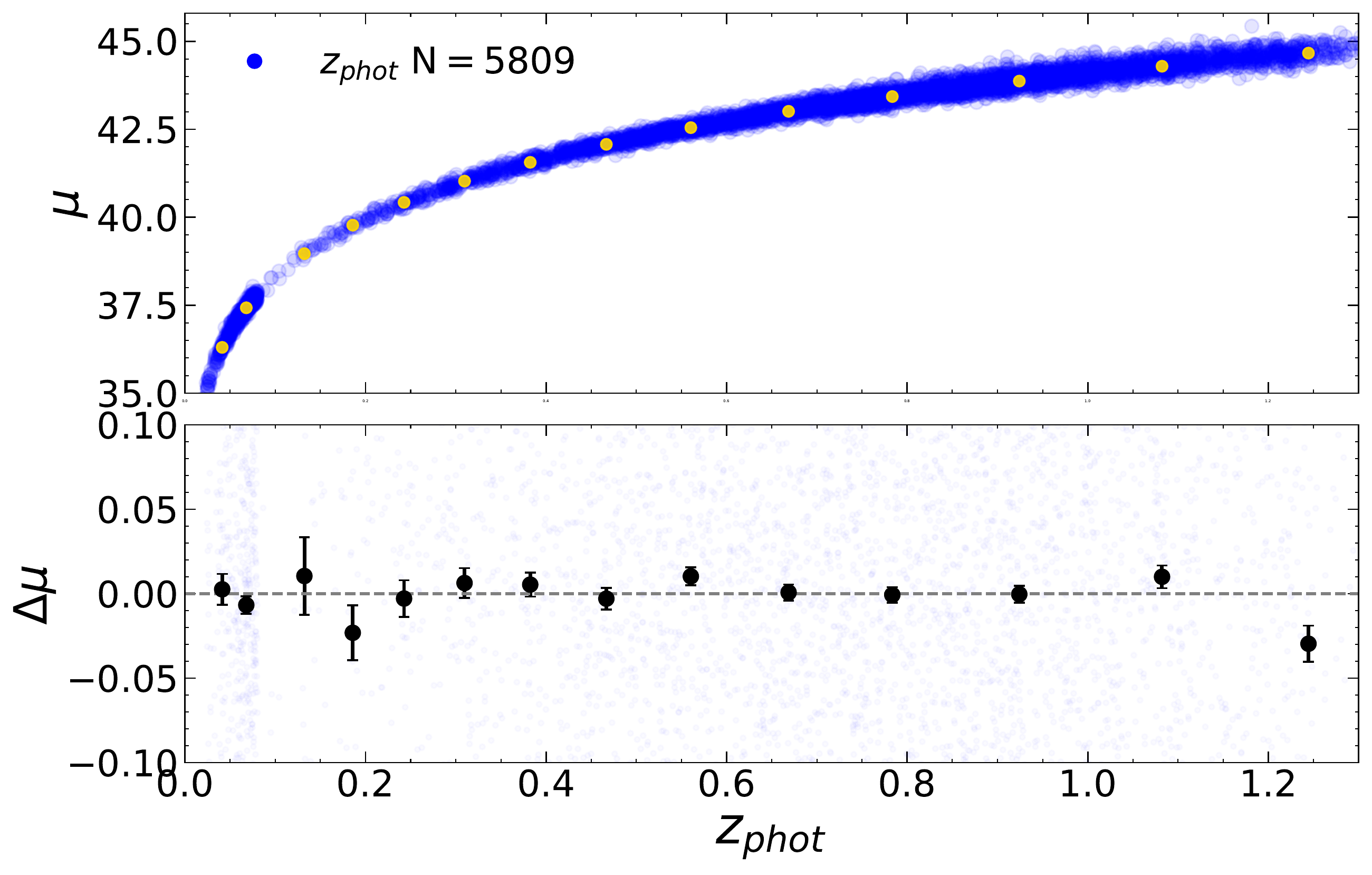}%
    \label{}
    }
   \caption{For one of the \NSAMPLE\ statistically independent samples, redshift binned (solid black circles) and unbinned Hubble diagram from BBC fit,
   for $\zspec$ (top) and $\zphot$ (bottom) samples.
   Each lower panel shows Hubble residual $\Delta\mu$ with respect to the true cosmology: the error bar shows the rms in each BBC redshift bin (same redshift bins as in Fig.~\ref{fig:musig_compare}).
    } 
   \label{fig:HD}
\end{figure}

For the subsections below, we define $w$-bias to be   $w - w_{\rm true}$ where $w$ is from the \wCDM\  cosmology fit. 
A similar definition is used for $w_0$ and $w_a$ for the \wwCDM\ model.

\subsection{ \wCDM Results}
For the \wCDM\ cosmology fits, Table~\ref{tb:wcdm_results} shows the average $w$-bias and average uncertainty among the $\NSAMPLE$ samples. 
The average $w$-bias is consistent with zero for both the $\zspec$ and $\zphot$ samples,
and also with and without systematic uncertainties. 
The $w$-bias precision is ${\sim}0.002$. 
The average $w$-uncertainty ($\AVGwsig$) for the $\zphot$ sample is \AVGwsigbiaszphotsyst, 
with systematics, and is only slightly improved compared to $\AVGwsig=$\AVGwsigbiaszspecsyst\ for the 
$\zspec$ sample. The  additional sensitivity from the host-galaxy $\zphot$ sample is small because the increased statistics are at higher redshifts where the  dark energy density fraction is much smaller compared to lower redshifts where the sample is  dominated by  spectroscopic redshifts.

\begin{table}[ht!]
\begin{center}
\caption{Summary of \wCDM\ Cosmology Fits}
\begin{tabular}{|cl|c|c|c|}
\hline
& redshift &              &           &   \\
& source    & Systematics 
            & $\AVGwbias$\tablenote{Average bias among $\NSAMPLE$ samples with uncertainty of std/$\sqrt{\NSAMPLE}$} 
            & $\AVGwsig$\tablenote{Average fitted uncertainty among $\NSAMPLE$ samples.}
               \\
[0.5ex]
 \hline
      & $\zspec$ & Stat only   & $-0.0008\pm 0.0020$  & $0.020$    \\
      &          & Stat$+$Syst & $-0.0027\pm 0.0025$  & \AVGwsigbiaszspecsyst     \\
  \hline      
     & $\zphot$ & Stat only   & $-0.0003\pm 0.0017$   & $0.020$    \\
     &          & Stat$+$Syst & $-0.0009\pm 0.0018$   & \AVGwsigbiaszphotsyst   \\
   \hline  
\end{tabular}
\label{tb:wcdm_results}
\end{center}
\end{table}
    

\subsection{ \wwCDM Results}

\begin{figure}
    \includegraphics[width=\columnwidth]{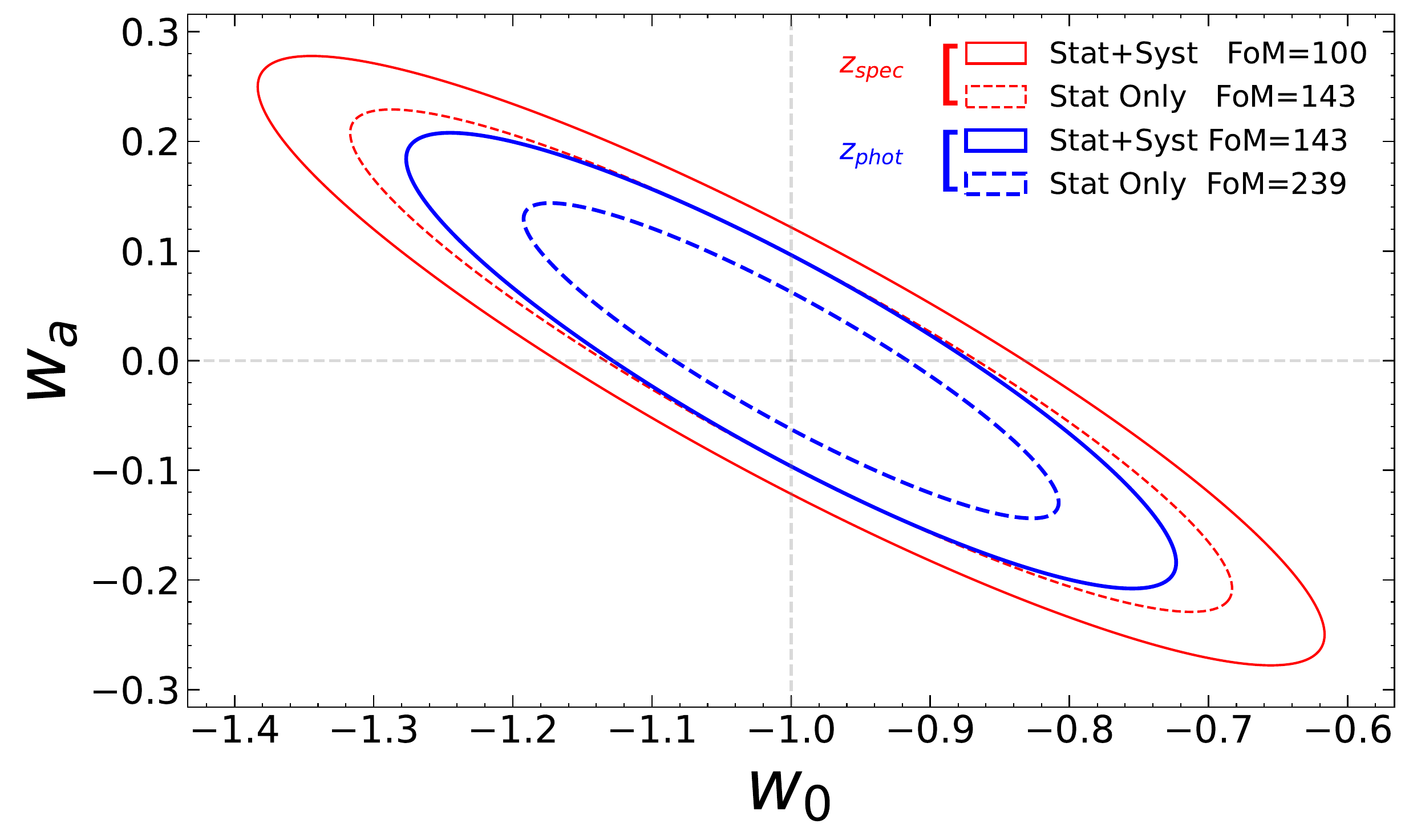}
    \caption{
    \ww $2\sigma$ $(95\%$ confidence$)$ contours and FoM  for a 
    single SNIa data sample combined with CMB prior,
    and shifted to be centered at $w_0,w_a=-1,0$.
    Contours for $\zphot$ ($\zspec$) are shown in blue (red). Solid (dashed) contours show stat+syst (stat-only).
    }
    \label{fig:w0wa_contours}
\end{figure}

For the \wwCDM\ model, the average bias, uncertainty, and FoM are shown in
Table~\ref{tb:w0wa_results}. 
While there was little improvement using the
$\zphot$ sample with the \wCDM\ model, the \wwCDM\  improvement is much more significant because higher redshift events, which are enhanced by the $\zphot$ sample,  are more sensitive to evolving dark energy ($w_a$).
With systematics, $\AVGFoM=$\AVGFOMzspecsyst\ for the $\zspec$ sample and
$\AVGFoM=$\AVGFOMzphotsyst\ for the $\zphot$ sample.
The \ww\ constraining power is shown in Fig.~\ref{fig:w0wa_contours} for a single
simulated data sample.

\begin{table*}[ht!]
\begin{center}
\caption{Summary of \wwCDM\ Cosmology Fits}
\begin{tabular}{|cl|c|c|c|c|c|c|}
\hline
& $z$ &              &     &      &  & & \\
& source    & Syst 
            & $\AVGwwbias$\tablenote{Average bias among $\NSAMPLE$ samples with uncertainty of std/$\sqrt{\NSAMPLE}$} 
            & $\AVGwabias$ & $\AVGwwsig$\tablenote{Average fitted uncertainty among $\NSAMPLE$ samples.} & $\AVGwasig$
            & $\AVGFoM$   \\
[0.5ex]
 \hline
      & $\zspec$ & Stat only   & $0.0083\pm0.0143$ &$-0.0658\pm0.0674$ & $0.076$ &$0.353$ & $136$\\
      &          & Stat$+$Syst & $0.0067\pm$\AVGwbiaszspecsyst &$-0.0683\pm$\AVGwabiaszspecsyst & $0.092$ &$0.418$ & \AVGFOMzspecsyst \\
  \hline      
     & $\zphot$ & Stat only   & $0.0029\pm0.0082$  & $-0.0228\pm 0.0342$ & $0.048$ & $0.211$ & \AVGFOMzphotstat\\
     &          & Stat$+$Syst & $0.0011\pm$\AVGwbiaszphotsyst  & $-0.0202\pm$\AVGwabiaszphotsyst & $0.071$ & $0.294$ & \AVGFOMzphotsyst \\
   \hline  
\end{tabular}
\label{tb:w0wa_results}
\end{center}
\end{table*}

\begin{figure}
    \includegraphics[width=\columnwidth]{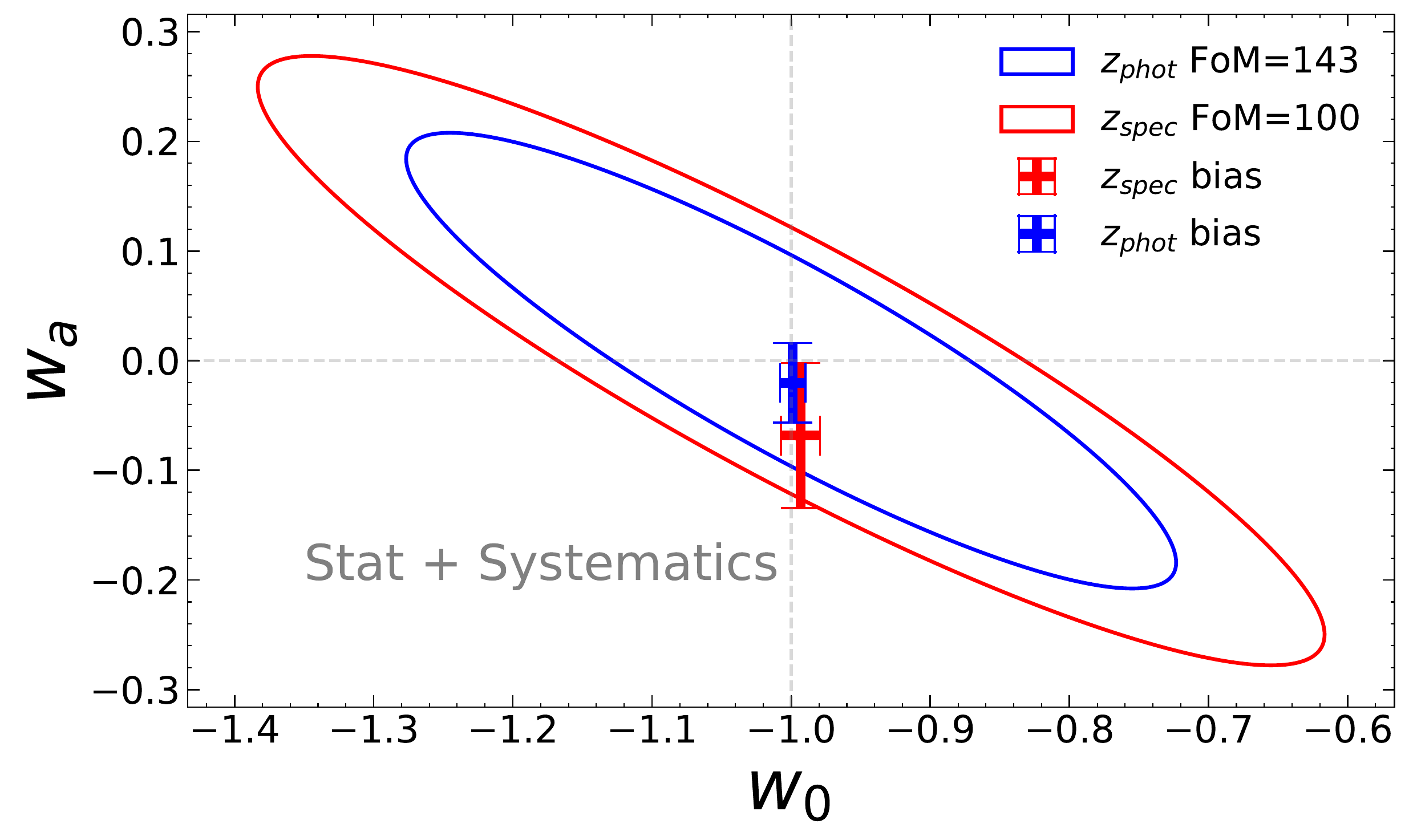}
    \caption{
\ww\ $2\sigma$ contours and FoM for $\zphot$ (blue) and $\zspec$ (red) for a single data sample.
    The crosses show the $\pm 1\sigma$ bias from averaging results over the \NSAMPLE\ simulated data samples.}
    \label{fig:w0wa_contours+bias}
\end{figure}

The average bias is consistent with zero for both $w_0$ and $w_a$.
For the $\zspec$ sample, the bias precision is ${\sim}0.015$ and ${\sim}0.07$ for $w_0$ and $w_a$, respectively. 
For the $\zphot$ sample, the bias precision is improved to 
${\sim}0.009$ and ${\sim}0.04$.
The \ww\ average bias is shown in Fig.~\ref{fig:w0wa_contours+bias}, 
and compared to the the \ww\ contours (statistical+systematic)
for a single sample.

For the $\zphot$ sample, the figure-of-merit averaged over \NSAMPLE\ samples is 
$\AVGFoM=$\AVGFOMzphotstat \ with only statistical uncertainties,
and drops to $\AVGFoM=$\AVGFOMzphotsyst \  when systematic uncertainties are included.
Since there are many systematics contributing to the decrease in $\AVGFoM$,
we quantify the impact of each systematic ``$i$'' by 
recomputing the covariance matrix separately for each systematic ($\COVsysti$), 
and repeating the cosmology fit for each $\COVsysti$.
We finally compute the FoM ratios
\begin{equation}
    \RatioFoM = \FoMSysti/\FoMStat~,
\end{equation}
where $\FoMSysti$ is the FoM from including only systematic $i$,
and $\FoMStat$ is the FoM without systematic uncertainties.
Note that $\RatioFoM \leq 1$.
Table~\ref{tb:syst_breakdown} shows the $\RatioFoM$, and the FoM degradation is dominated
by the calibration systematics. 

\subsection{Discussion of photo-\protect{$z$} Systematics}

\begin{figure}[h]
    \centering
    \subfloat{{\includegraphics[width=.418\textwidth]{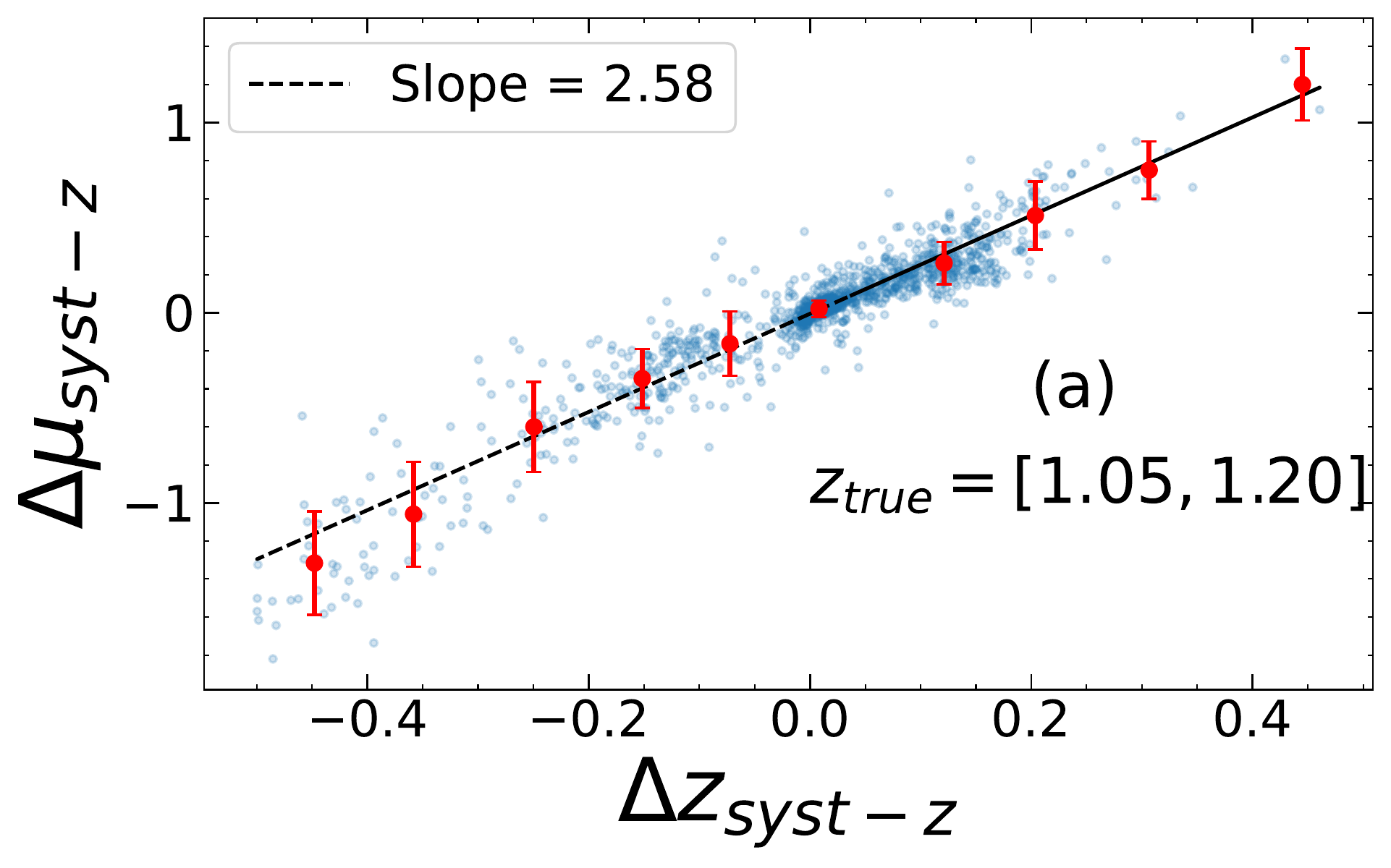} }}

    \subfloat{{\includegraphics[width=.420\textwidth]{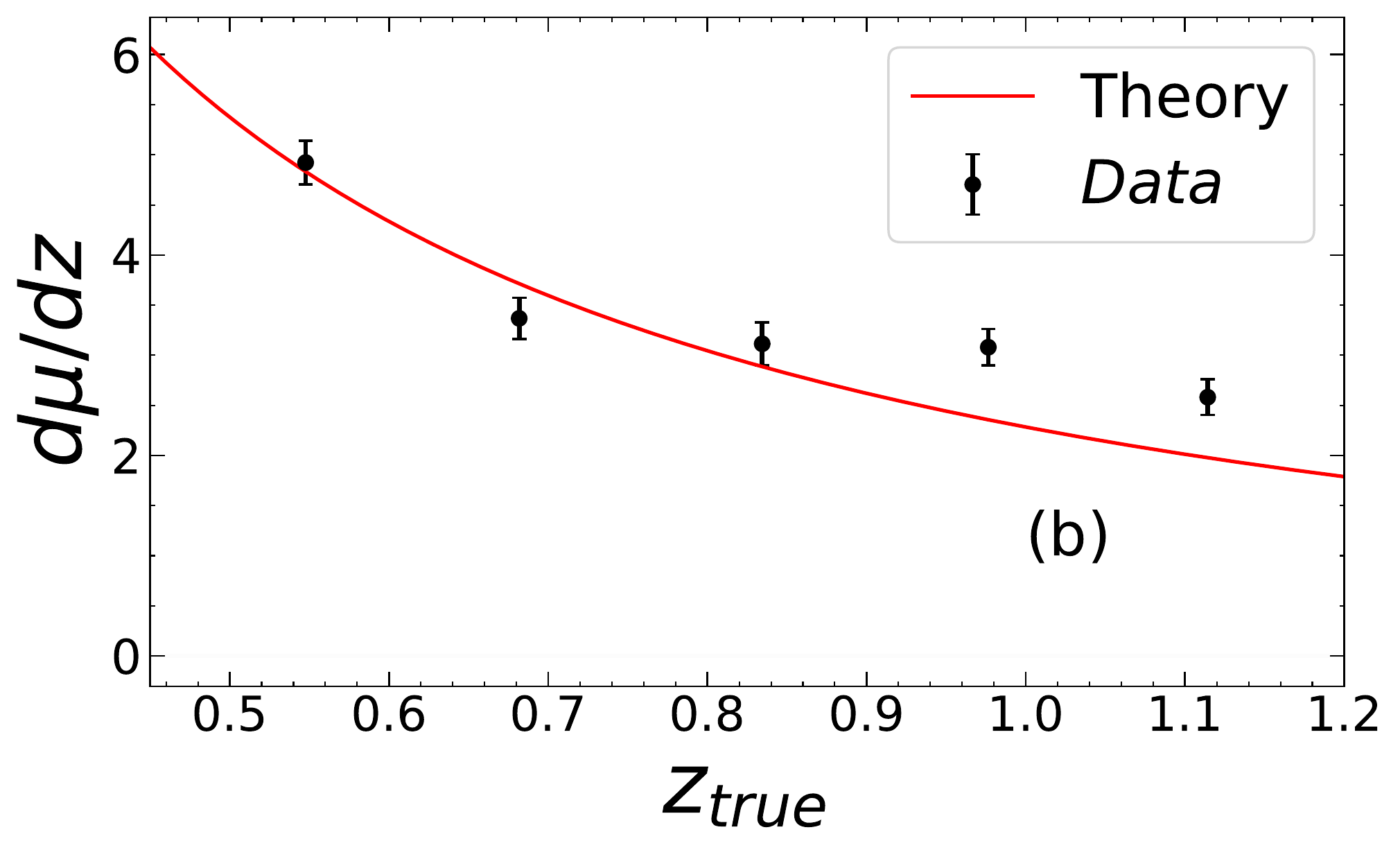} }}
     \caption{
     (a) $\dmusyst$ vs. $\dzsyst$ for 0.01 systematic shift in host-galaxy photo-$z$,
        and linear fit with slope $d\mu/dz$,
        (b) fitted slope $d\mu/dz$ in five $\ztrue$ bins (black circles with error bars),
        and $\LCDM$ theory curve in red.
     }
    \label{fig:dmudz}
\end{figure}

The 0.01 photo-$z$ shift systematic has a small (2\%) effect on FoM
for three reasons. First, the combined SN+host light curve fit
results in an average fitted redshift error of ${\sim}0.004$, or about half the host photo-$z$ error.
Second, this photo-$z$ systematic does not affect $\zspec$ events which dominate the 
lower redshift region below about 0.5 (Fig.~\ref{fig:musig_compare}a),
and this $\zspec$ region is most sensitive to redshift errors.
The final reason is that the fitted $\zphot$ and SALT2 color are anti-correlated and thus a 
larger (smaller) $\zphot$ results in bluer (redder) color,
and this change in color self-corrects the distance error as illustrated in
Fig.~\ref{fig:dmudz}.

To describe this distance self-correction,
we first define $\dzsyst$ as the difference between SALT2-fitted
$\zphot$ with 0.01 host-galaxy photo-$z$ shift and nominal photo-$z$, and similarly
define $\dmusyst$ as the distance difference from Eq.~\ref{eq:mu}. 
Fig~\ref{fig:dmudz}a shows $\dmusyst$ vs $\dzsyst$ and a linear fit
for the slope, $d\mu/dz$, in one of five $\ztrue$ bins. 
Fig~\ref{fig:dmudz}b shows the measured $d\mu/dz$ slope in five $\ztrue$ bins (black circles) along
with the $\LCDM$ theory curve in red.
In the ideal limit where the measured $d\mu/dz$ exactly equal theory $d\mu/dz$, 
the distance self-correction is perfect and results in no systematic uncertainty.
Here the measured $d\mu/dz$ are close to the theory curve, and thus the distance
error is mostly corrected.

\begin{table}
\caption{FoM-Ratio $\RatioFoM$ for Each Systematic (\wwCDM \ model)}
\begin{center}
\begin{tabular}{| l | c |} 
 \hline
                &  $\RatioFoM$  \\ 
  Systematic(s) &  $\zphot$\ \ \ $\zspec$\ \ \ \\  [0.5ex] 
 \hline
  None (stat only) & $1.00\ \  1.00\ \  $ \\
 \hline
  MWEBV  & $0.99\ \  1.00\ \  $  \\
 \hline
 ZERRSCALE & $0.99\ \  1.00\ \  $  \\
 \hline
 zSHIFT  &   $0.98\ \  0.99\ \  $  \\ 
 \hline
 
 Photo-z Shift & $0.98\ \  0.99\ \  $  \\
 \hline
  CAL$\_$WAVE & $0.90\ \  0.94\ \  $ \\
\hline
CAL$\_Z_p$ & $0.71\ \  0.75\ \  $\\
 \hline
 
CAL&$0.64\ \  0.71\ \  $  \\  
 \hline\hline
 Stat $+$ All Syst  & $0.61\ \  0.70\ \  $ \\ 
 \hline
\end{tabular}
\label{tb:syst_breakdown}
\end{center}
\end{table}

 To gain further insight into the photo-$z$ sensitivity, we first consider a naive systematic
of shifting the fitted $\zphot$ by 0.01 after the light curve fit, for the subset without a $\zspec$.
In this test, the compensating $d\mu/dz$ points in Fig.~\ref{fig:dmudz} are forced to be zero,
and there is no systematic reduction from a combined SN+host fit.
Fitting the \wwCDM\ model without $\COVsyst$, 
the $w_0$ and $w_a$ biases are 0.03 and 0.15, respectively.
Next we consider the realistic case of shifting the host photo-$z$ before the SALT2 light curve fit;
the corresponding $w_0$ and $w_a$ biases are 0.001 and 0.003, 
more than an order of magnitude smaller
than the naive systematic. 
While we have included an explicit host-galaxy photo-$z$ systematic, there is no explicit analogue 
for the SN.
The SN photo-$z$ systematic is accounted for by the calibration and Galactic extinction contributions 
to the systematic uncertainty budget in Table~\ref{tb:syst_sources},
but it is difficult to untangle the impact of these systematics on distance and photo-$z$.

\section{Conclusions}\label{sec:conclude} 
In this work we presented cosmological dark energy constraints
for simulated \acro-SN~Ia data,
and we continued the development of publicly available codes from \SNANA\ and \pippin
to analyse the the data
with a host galaxy photo-$z$ prior.
For the \wwCDM\ model, the dark energy figure of merit is FoM${\sim}$\AVGFOMzphotstat\ 
with only statistical uncertainties, and drops to ${\sim}$\AVGFOMzphotsyst\  
with systematic uncertainties (Fig.~\ref{fig:w0wa_contours}).
This $\zphot$ FoM is 50\% larger than the FoM obtained from the
$\zspec$ subset that has a \spec\ redshift from the host or SN.
Averaging \NSAMPLE\ independent data samples,  
the average bias on $w_0$ and $w_a$ is consistent with zero.

The systematic uncertainty from the host-galaxy photo-$z$ results in only a 2\% reduction in the FoM. 
This small impact is due to
i) nearly complete $\zspec$ at lower redshifts, 
ii) smaller $\zphot$ bias from combining the SN and host, and
iii) anti-correlations between redshift and color that greatly reduce the distance error.
While good $\zspec$ coverage is feasible for the DDF, the WFD will likely have
less $\zspec$ coverage and using host-galaxy photo-$z$'s at lower redshifts may increase 
the systematic uncertainty compared to this DDF analysis.

 Simulated projections tend to be overly optimistic before a survey begins, particularly for the depth and  average PSF. 
 However, there are three key factors that are likely to improve future results: 
1) here we simulated only 30\% of the 10-year baseline survey,
2) we used a CMB prior with constraining power to match \citet{Planck2018},
 and did not assume improved CMB constraints during the LSST era,
3) we did not include the ${\sim}50$\% FoM-increase from 
fitting an unbinned Hubble diagram;
this improvement awaits resolving the large \ww\ bias associated with unbinned results.

Most SNIa-cosmology analyses over the past decade have used redshift binned Hubble diagrams. 
These analyses include, 
JLA \citep{Betoule2014},
Pantheon \citep{Scolnic2018},
PS1 single instrument \citep{Jones2018} and
 DES \citep{DES3YR}. 
 The recent demonstration of smaller uncertainties with an unbinned Hubble diagram has not been 
 rigorously tested until our analysis that shows biased cosmology parameters. 
 We therefore encourage community effort to resolve this issue.

The next major effort  is to develop the cosmology analysis for 
samples that include  non-SNIa contamination, host galaxy mis-association,
and a more complete list of systematic uncertainties that includes host galaxy photo-$z$ model and intrinsic scatter of the SN brightness. 
Cosmology analyses using photometric classification and \spec\ redshifts 
have been well developed on real data from PS1 \citep{Jones2018}
and from DES \citep{Vincenzi2022}. Here we have developed and
demonstrated a complimentary analysis using photometric redshifts 
and a \specy\ confirmed sample.

\acknowledgments 
Author contributions are listed below. \\
A.~Mitra: co-lead project, SNANA simulations and analysis, writing \\
R.~Kessler: co-lead project, software, analysis, writing \\
S.~More: writing, review \\
R.~Hlozek: development of PLAsTiCC challenge. \\
\medskip

AM acknowledges the funding of the Science Committee of the Ministry of Education and Science of the Republic of Kazakhstan 
(Grant No. AP08856149) and Nazarbayev University Faculty Development Competitive Research Grant Program No 11022021FD2912. 
RK acknowledges pipeline scientist support from the LSST Dark Energy Science Collaboration.
This work was completed in part with resources provided by the University of Chicago’s Research Computing Center.

This paper has passed an internal review by the DESC  and we thank the DESC internal reviewers: Dan Scolnic, Bruno Sanchez and Martine Lokken. 

The DESC acknowledges ongoing support from the Institut National de 
Physique Nucl\'eaire et de Physique des Particules in France; the 
Science \& Technology Facilities Council in the United Kingdom; and the
Department of Energy, the National Science Foundation, and the LSST 
Corporation in the United States.  DESC uses resources of the IN2P3 
Computing Center (CC-IN2P3--Lyon/Villeurbanne - France) funded by the 
Centre National de la Recherche Scientifique; the National Energy 
Research Scientific Computing Center, a DOE Office of Science User 
Facility supported by the Office of Science of the U.S.\ Department of
Energy under Contract No.\ DE-AC02-05CH11231; STFC DiRAC HPC Facilities, 
funded by UK BIS National E-infrastructure capital grants; and the UK 
particle physics grid, supported by the GridPP Collaboration.  This 
work was performed in part under DOE Contract DE-AC02-76SF00515.

\bibliographystyle{apsrev}
\bibliography{main}

\begin{thebibliography}{59}
\expandafter\ifx\csname natexlab\endcsname\relax\def\natexlab#1{#1}\fi
\expandafter\ifx\csname bibnamefont\endcsname\relax
  \def\bibnamefont#1{#1}\fi
\expandafter\ifx\csname bibfnamefont\endcsname\relax
  \def\bibfnamefont#1{#1}\fi
\expandafter\ifx\csname citenamefont\endcsname\relax
  \def\citenamefont#1{#1}\fi
\expandafter\ifx\csname url\endcsname\relax
  \def\url#1{\texttt{#1}}\fi
\expandafter\ifx\csname urlprefix\endcsname\relax\def\urlprefix{URL }\fi
\providecommand{\bibinfo}[2]{#2}
\providecommand{\eprint}[2][]{\url{#2}}

\bibitem[{\citenamefont{{Perlmutter} et~al.}(1999)\citenamefont{{Perlmutter},
  {Aldering}, {Goldhaber}, {Knop}, {Nugent} et~al.}}]{perl}
\bibinfo{author}{\bibfnamefont{S.}~\bibnamefont{{Perlmutter}}},
  \bibinfo{author}{\bibfnamefont{G.}~\bibnamefont{{Aldering}}},
  \bibinfo{author}{\bibfnamefont{G.}~\bibnamefont{{Goldhaber}}},
  \bibinfo{author}{\bibfnamefont{R.~A.} \bibnamefont{{Knop}}},
  \bibinfo{author}{\bibfnamefont{P.}~\bibnamefont{{Nugent}}},
  \bibnamefont{et~al.}, \bibinfo{journal}{Astrophys. J.}
  \textbf{\bibinfo{volume}{517}}, \bibinfo{pages}{565} (\bibinfo{year}{1999}),
  \eprint{astro-ph/9812133}.

\bibitem[{\citenamefont{{Riess} et~al.}(1998)\citenamefont{{Riess},
  {Filippenko}, {Challis}, {Clocchiatti}, {Diercks} et~al.}}]{adam}
\bibinfo{author}{\bibfnamefont{A.~G.} \bibnamefont{{Riess}}},
  \bibinfo{author}{\bibfnamefont{A.~V.} \bibnamefont{{Filippenko}}},
  \bibinfo{author}{\bibfnamefont{P.}~\bibnamefont{{Challis}}},
  \bibinfo{author}{\bibfnamefont{A.}~\bibnamefont{{Clocchiatti}}},
  \bibinfo{author}{\bibfnamefont{A.}~\bibnamefont{{Diercks}}},
  \bibnamefont{et~al.}, \bibinfo{journal}{Astron. J}
  \textbf{\bibinfo{volume}{116}}, \bibinfo{pages}{1009} (\bibinfo{year}{1998}),
  \eprint{astro-ph/9805201}.

\bibitem[{\citenamefont{{Linder}}(2003)}]{cpl}
\bibinfo{author}{\bibfnamefont{E.~V.} \bibnamefont{{Linder}}},
  \bibinfo{journal}{Phys. Rev. Lett.} \textbf{\bibinfo{volume}{90}},
  \bibinfo{eid}{091301} (\bibinfo{year}{2003}), \eprint{astro-ph/0208512}.

\bibitem[{\citenamefont{{Chevallier} and {Polarski}}(2001)}]{cpl2}
\bibinfo{author}{\bibfnamefont{M.}~\bibnamefont{{Chevallier}}}
  \bibnamefont{and}
  \bibinfo{author}{\bibfnamefont{D.}~\bibnamefont{{Polarski}}},
  \bibinfo{journal}{International Journal of Modern Physics D}
  \textbf{\bibinfo{volume}{10}}, \bibinfo{pages}{213} (\bibinfo{year}{2001}),
  \eprint{gr-qc/0009008}.

\bibitem[{\citenamefont{{Betoule} et~al.}(2014)}]{Betoule2014}
\bibinfo{author}{\bibfnamefont{M.}~\bibnamefont{{Betoule}}}
  \bibnamefont{et~al.}, \bibinfo{journal}{Astron. Astrophys.}
  \textbf{\bibinfo{volume}{568}}, \bibinfo{eid}{A22} (\bibinfo{year}{2014}),
  \eprint{1401.4064}.

\bibitem[{\citenamefont{{Scolnic} et~al.}(2018)\citenamefont{{Scolnic},
  {Jones}, {Rest}, {Pan}, {Chornock}, {Foley}, {Huber}, {Kessler}, {Narayan},
  {Riess} et~al.}}]{Scolnic2018}
\bibinfo{author}{\bibfnamefont{D.~M.} \bibnamefont{{Scolnic}}},
  \bibinfo{author}{\bibfnamefont{D.~O.} \bibnamefont{{Jones}}},
  \bibinfo{author}{\bibfnamefont{A.}~\bibnamefont{{Rest}}},
  \bibinfo{author}{\bibfnamefont{Y.~C.} \bibnamefont{{Pan}}},
  \bibinfo{author}{\bibfnamefont{R.}~\bibnamefont{{Chornock}}},
  \bibinfo{author}{\bibfnamefont{R.~J.} \bibnamefont{{Foley}}},
  \bibinfo{author}{\bibfnamefont{M.~E.} \bibnamefont{{Huber}}},
  \bibinfo{author}{\bibfnamefont{R.}~\bibnamefont{{Kessler}}},
  \bibinfo{author}{\bibfnamefont{G.}~\bibnamefont{{Narayan}}},
  \bibinfo{author}{\bibfnamefont{A.~G.} \bibnamefont{{Riess}}},
  \bibnamefont{et~al.}, \bibinfo{journal}{Astrophys. J.}
  \textbf{\bibinfo{volume}{859}}, \bibinfo{eid}{101} (\bibinfo{year}{2018}),
  \eprint{1710.00845}.

\bibitem[{\citenamefont{{Abbott} et~al.}(2019)\citenamefont{{Abbott}, {Allam},
  {Andersen}, {Angus}, {Asorey} et~al.}}]{DES3YR}
\bibinfo{author}{\bibfnamefont{T.~M.~C.} \bibnamefont{{Abbott}}},
  \bibinfo{author}{\bibfnamefont{S.}~\bibnamefont{{Allam}}},
  \bibinfo{author}{\bibfnamefont{P.}~\bibnamefont{{Andersen}}},
  \bibinfo{author}{\bibfnamefont{C.}~\bibnamefont{{Angus}}},
  \bibinfo{author}{\bibfnamefont{J.}~\bibnamefont{{Asorey}}},
  \bibnamefont{et~al.} (\bibinfo{collaboration}{DES Collaboration}),
  \bibinfo{journal}{Astrophys. J. Lett.} \textbf{\bibinfo{volume}{872}},
  \bibinfo{eid}{L30} (\bibinfo{year}{2019}), \eprint{1811.02374}.

\bibitem[{\citenamefont{Brout et~al.}(2022)}]{pantheon_new}
\bibinfo{author}{\bibfnamefont{D.}~\bibnamefont{Brout}} \bibnamefont{et~al.}
  (\bibinfo{year}{2022}), \eprint{2202.04077}.

\bibitem[{\citenamefont{Lochner et~al.}(2016)}]{Lochner2016}
\bibinfo{author}{\bibfnamefont{M.}~\bibnamefont{Lochner}} \bibnamefont{et~al.},
  \bibinfo{journal}{Astrophys. J. Suppl.} \textbf{\bibinfo{volume}{225}},
  \bibinfo{pages}{31} (\bibinfo{year}{2016}), \eprint{1603.00882}.

\bibitem[{\citenamefont{{M{\"o}ller} and {de
  Boissi{\`e}re}}(2020)}]{Moller2020}
\bibinfo{author}{\bibfnamefont{A.}~\bibnamefont{{M{\"o}ller}}}
  \bibnamefont{and} \bibinfo{author}{\bibfnamefont{T.}~\bibnamefont{{de
  Boissi{\`e}re}}}, \bibinfo{journal}{Mon. Not. Roy. Astron. Soc.}
  \textbf{\bibinfo{volume}{491}}, \bibinfo{pages}{4277} (\bibinfo{year}{2020}),
  \eprint{1901.06384}.

\bibitem[{\citenamefont{{Kessler} et~al.}(2010)\citenamefont{{Kessler},
  {Cinabro}, {Bassett}, {Dilday}, {Frieman} et~al.}}]{Kessler2010}
\bibinfo{author}{\bibfnamefont{R.}~\bibnamefont{{Kessler}}},
  \bibinfo{author}{\bibfnamefont{D.}~\bibnamefont{{Cinabro}}},
  \bibinfo{author}{\bibfnamefont{B.}~\bibnamefont{{Bassett}}},
  \bibinfo{author}{\bibfnamefont{B.}~\bibnamefont{{Dilday}}},
  \bibinfo{author}{\bibfnamefont{J.~A.} \bibnamefont{{Frieman}}},
  \bibnamefont{et~al.}, \bibinfo{journal}{Astrophys. J.}
  \textbf{\bibinfo{volume}{717}}, \bibinfo{pages}{40} (\bibinfo{year}{2010}),
  \eprint{1001.0738}.

\bibitem[{\citenamefont{{Palanque-Delabrouille}
  et~al.}(2010)\citenamefont{{Palanque-Delabrouille}, {Ruhlmann-Kleider},
  {Pascal}, {Rich}, {Guy}, {Bazin}, {Astier}, {Balland}, {Basa}, {Carlberg}
  et~al.}}]{Palanque2010}
\bibinfo{author}{\bibfnamefont{N.}~\bibnamefont{{Palanque-Delabrouille}}},
  \bibinfo{author}{\bibfnamefont{V.}~\bibnamefont{{Ruhlmann-Kleider}}},
  \bibinfo{author}{\bibfnamefont{S.}~\bibnamefont{{Pascal}}},
  \bibinfo{author}{\bibfnamefont{J.}~\bibnamefont{{Rich}}},
  \bibinfo{author}{\bibfnamefont{J.}~\bibnamefont{{Guy}}},
  \bibinfo{author}{\bibfnamefont{G.}~\bibnamefont{{Bazin}}},
  \bibinfo{author}{\bibfnamefont{P.}~\bibnamefont{{Astier}}},
  \bibinfo{author}{\bibfnamefont{C.}~\bibnamefont{{Balland}}},
  \bibinfo{author}{\bibfnamefont{S.}~\bibnamefont{{Basa}}},
  \bibinfo{author}{\bibfnamefont{R.~G.} \bibnamefont{{Carlberg}}},
  \bibnamefont{et~al.}, \bibinfo{journal}{Astron. Astrophys.}
  \textbf{\bibinfo{volume}{514}}, \bibinfo{eid}{A63} (\bibinfo{year}{2010}),
  \eprint{0911.1629}.

\bibitem[{\citenamefont{Roberts et~al.}(2017)\citenamefont{Roberts, Lochner,
  Fonseca, Bassett, Lablanche, and Agarwal}}]{zbeams}
\bibinfo{author}{\bibfnamefont{E.}~\bibnamefont{Roberts}},
  \bibinfo{author}{\bibfnamefont{M.}~\bibnamefont{Lochner}},
  \bibinfo{author}{\bibfnamefont{J.}~\bibnamefont{Fonseca}},
  \bibinfo{author}{\bibfnamefont{B.~A.} \bibnamefont{Bassett}},
  \bibinfo{author}{\bibfnamefont{P.-Y.} \bibnamefont{Lablanche}},
  \bibnamefont{and} \bibinfo{author}{\bibfnamefont{S.}~\bibnamefont{Agarwal}},
  \bibinfo{journal}{J. Cosmol. Astropart. Phys.} \textbf{\bibinfo{volume}{10}},
  \bibinfo{pages}{036} (\bibinfo{year}{2017}), \eprint{1704.07830}.

\bibitem[{\citenamefont{{Kunz} et~al.}(2007)\citenamefont{{Kunz}, {Bassett},
  and {Hlozek}}}]{kunz}
\bibinfo{author}{\bibfnamefont{M.}~\bibnamefont{{Kunz}}},
  \bibinfo{author}{\bibfnamefont{B.~A.} \bibnamefont{{Bassett}}},
  \bibnamefont{and} \bibinfo{author}{\bibfnamefont{R.~A.}
  \bibnamefont{{Hlozek}}}, \bibinfo{journal}{\prd}
  \textbf{\bibinfo{volume}{75}}, \bibinfo{eid}{103508} (\bibinfo{year}{2007}),
  \eprint{astro-ph/0611004}.

\bibitem[{\citenamefont{{Hlozek} et~al.}(2012)\citenamefont{{Hlozek}, {Kunz},
  {Bassett}, {Smith}, {Newling} et~al.}}]{hlozek12}
\bibinfo{author}{\bibfnamefont{R.}~\bibnamefont{{Hlozek}}},
  \bibinfo{author}{\bibfnamefont{M.}~\bibnamefont{{Kunz}}},
  \bibinfo{author}{\bibfnamefont{B.}~\bibnamefont{{Bassett}}},
  \bibinfo{author}{\bibfnamefont{M.}~\bibnamefont{{Smith}}},
  \bibinfo{author}{\bibfnamefont{J.}~\bibnamefont{{Newling}}},
  \bibnamefont{et~al.}, \bibinfo{journal}{Astrophys. J.}
  \textbf{\bibinfo{volume}{752}}, \bibinfo{eid}{79} (\bibinfo{year}{2012}),
  \eprint{1111.5328}.

\bibitem[{\citenamefont{{Jones} et~al.}(2018)\citenamefont{{Jones}, {Scolnic},
  {Riess}, {Rest}, {Kirshner} et~al.}}]{Jones2018}
\bibinfo{author}{\bibfnamefont{D.~O.} \bibnamefont{{Jones}}},
  \bibinfo{author}{\bibfnamefont{D.~M.} \bibnamefont{{Scolnic}}},
  \bibinfo{author}{\bibfnamefont{A.~G.} \bibnamefont{{Riess}}},
  \bibinfo{author}{\bibfnamefont{A.}~\bibnamefont{{Rest}}},
  \bibinfo{author}{\bibfnamefont{R.}~\bibnamefont{{Kirshner}}},
  \bibnamefont{et~al.}, \bibinfo{journal}{Astrophys. J.}
  \textbf{\bibinfo{volume}{857}}, \bibinfo{eid}{51} (\bibinfo{year}{2018}),
  \eprint{1710.00846}.

\bibitem[{\citenamefont{{Kessler} and {Scolnic}}(2017)}]{bbc}
\bibinfo{author}{\bibfnamefont{R.}~\bibnamefont{{Kessler}}} \bibnamefont{and}
  \bibinfo{author}{\bibfnamefont{D.}~\bibnamefont{{Scolnic}}},
  \bibinfo{journal}{Astrophys. J.} \textbf{\bibinfo{volume}{836}},
  \bibinfo{eid}{56} (\bibinfo{year}{2017}), \eprint{1610.04677}.

\bibitem[{\citenamefont{{Vincenzi} et~al.}(2022)\citenamefont{{Vincenzi},
  {Sullivan}, {M{\"o}ller}, {Armstrong}, {Bassett}, {Brout}, {Carollo}, {Carr},
  {Davis}, {Frohmaier} et~al.}}]{Vincenzi2022}
\bibinfo{author}{\bibfnamefont{M.}~\bibnamefont{{Vincenzi}}},
  \bibinfo{author}{\bibfnamefont{M.}~\bibnamefont{{Sullivan}}},
  \bibinfo{author}{\bibfnamefont{A.}~\bibnamefont{{M{\"o}ller}}},
  \bibinfo{author}{\bibfnamefont{P.}~\bibnamefont{{Armstrong}}},
  \bibinfo{author}{\bibfnamefont{B.~A.} \bibnamefont{{Bassett}}},
  \bibinfo{author}{\bibfnamefont{D.}~\bibnamefont{{Brout}}},
  \bibinfo{author}{\bibfnamefont{D.}~\bibnamefont{{Carollo}}},
  \bibinfo{author}{\bibfnamefont{A.}~\bibnamefont{{Carr}}},
  \bibinfo{author}{\bibfnamefont{T.~M.} \bibnamefont{{Davis}}},
  \bibinfo{author}{\bibfnamefont{C.}~\bibnamefont{{Frohmaier}}},
  \bibnamefont{et~al.}, \bibinfo{journal}{"Mon. Not. Roy. Astron. Soc."}
  (\bibinfo{year}{2022}), \eprint{2111.10382}.

\bibitem[{\citenamefont{{Guy} et~al.}(2007)\citenamefont{{Guy}, {Astier},
  {Baumont}, {Hardin}, {Pain}, {Regnault}, {Basa}, {Carlberg}, {Conley},
  {Fabbro} et~al.}}]{salt2}
\bibinfo{author}{\bibfnamefont{J.}~\bibnamefont{{Guy}}},
  \bibinfo{author}{\bibfnamefont{P.}~\bibnamefont{{Astier}}},
  \bibinfo{author}{\bibfnamefont{S.}~\bibnamefont{{Baumont}}},
  \bibinfo{author}{\bibfnamefont{D.}~\bibnamefont{{Hardin}}},
  \bibinfo{author}{\bibfnamefont{R.}~\bibnamefont{{Pain}}},
  \bibinfo{author}{\bibfnamefont{N.}~\bibnamefont{{Regnault}}},
  \bibinfo{author}{\bibfnamefont{S.}~\bibnamefont{{Basa}}},
  \bibinfo{author}{\bibfnamefont{R.~G.} \bibnamefont{{Carlberg}}},
  \bibinfo{author}{\bibfnamefont{A.}~\bibnamefont{{Conley}}},
  \bibinfo{author}{\bibfnamefont{S.}~\bibnamefont{{Fabbro}}},
  \bibnamefont{et~al.}, \bibinfo{journal}{Astron. Astrophys.}
  \textbf{\bibinfo{volume}{466}}, \bibinfo{pages}{11} (\bibinfo{year}{2007}),
  \eprint{astro-ph/0701828}.

\bibitem[{\citenamefont{Dai et~al.}(2018)}]{Dai2018}
\bibinfo{author}{\bibfnamefont{M.}~\bibnamefont{Dai}} \bibnamefont{et~al.},
  \bibinfo{journal}{Mon. Not. Roy. Astron. Soc.}
  \textbf{\bibinfo{volume}{477}}, \bibinfo{pages}{4142} (\bibinfo{year}{2018}),
  \eprint{1701.05689}.

\bibitem[{\citenamefont{{Chen} et~al.}(2022)\citenamefont{{Chen}, {Scolnic},
  {Rozo}, {Rykoff}, {Popovic}, {Kessler}, {Vincenzi}, {Davis}, {Armstrong},
  {Brout} et~al.}}]{Chen2022}
\bibinfo{author}{\bibfnamefont{R.}~\bibnamefont{{Chen}}},
  \bibinfo{author}{\bibfnamefont{D.}~\bibnamefont{{Scolnic}}},
  \bibinfo{author}{\bibfnamefont{E.}~\bibnamefont{{Rozo}}},
  \bibinfo{author}{\bibfnamefont{E.~S.} \bibnamefont{{Rykoff}}},
  \bibinfo{author}{\bibfnamefont{B.}~\bibnamefont{{Popovic}}},
  \bibinfo{author}{\bibfnamefont{R.}~\bibnamefont{{Kessler}}},
  \bibinfo{author}{\bibfnamefont{M.}~\bibnamefont{{Vincenzi}}},
  \bibinfo{author}{\bibfnamefont{T.~M.} \bibnamefont{{Davis}}},
  \bibinfo{author}{\bibfnamefont{P.}~\bibnamefont{{Armstrong}}},
  \bibinfo{author}{\bibfnamefont{D.}~\bibnamefont{{Brout}}},
  \bibnamefont{et~al.}, \bibinfo{journal}{arXiv e-prints}
  \bibinfo{eid}{arXiv:2202.10480} (\bibinfo{year}{2022}), \eprint{2202.10480}.

\bibitem[{\citenamefont{{Linder} and {Mitra}}(2019)}]{Linder2019}
\bibinfo{author}{\bibfnamefont{E.~V.} \bibnamefont{{Linder}}} \bibnamefont{and}
  \bibinfo{author}{\bibfnamefont{A.}~\bibnamefont{{Mitra}}},
  \bibinfo{journal}{Phys. Rev. D} \textbf{\bibinfo{volume}{100}},
  \bibinfo{eid}{043542} (\bibinfo{year}{2019}), \eprint{1907.00985}.

\bibitem[{\citenamefont{Mitra and Linder}(2021)}]{Mitra2021}
\bibinfo{author}{\bibfnamefont{A.}~\bibnamefont{Mitra}} \bibnamefont{and}
  \bibinfo{author}{\bibfnamefont{E.~V.} \bibnamefont{Linder}},
  \bibinfo{journal}{Phys. Rev. D} \textbf{\bibinfo{volume}{103}},
  \bibinfo{pages}{023524} (\bibinfo{year}{2021}), \eprint{2011.08206}.

\bibitem[{\citenamefont{{Kessler}
  et~al.}(2019{\natexlab{a}})\citenamefont{{Kessler}, {Narayan}, {Avelino},
  {Bachelet}, {Biswas}, {Brown}, {Chernoff}, {Connolly}, {Dai}, {Daniel}
  et~al.}}]{plasticc_K2019}
\bibinfo{author}{\bibfnamefont{R.}~\bibnamefont{{Kessler}}},
  \bibinfo{author}{\bibfnamefont{G.}~\bibnamefont{{Narayan}}},
  \bibinfo{author}{\bibfnamefont{A.}~\bibnamefont{{Avelino}}},
  \bibinfo{author}{\bibfnamefont{E.}~\bibnamefont{{Bachelet}}},
  \bibinfo{author}{\bibfnamefont{R.}~\bibnamefont{{Biswas}}},
  \bibinfo{author}{\bibfnamefont{P.~J.} \bibnamefont{{Brown}}},
  \bibinfo{author}{\bibfnamefont{D.~F.} \bibnamefont{{Chernoff}}},
  \bibinfo{author}{\bibfnamefont{A.~J.} \bibnamefont{{Connolly}}},
  \bibinfo{author}{\bibfnamefont{M.}~\bibnamefont{{Dai}}},
  \bibinfo{author}{\bibfnamefont{S.}~\bibnamefont{{Daniel}}},
  \bibnamefont{et~al.}, \bibinfo{journal}{Publ. Astron. Soc. Pac.}
  \textbf{\bibinfo{volume}{131}}, \bibinfo{pages}{094501}
  (\bibinfo{year}{2019}{\natexlab{a}}), \eprint{1903.11756}.

\bibitem[{\citenamefont{{Kessler} et~al.}(2009)\citenamefont{{Kessler},
  {Bernstein}, {Cinabro}, {Dilday}, {Frieman}, {Jha}, {Kuhlmann}, {Miknaitis},
  {Sako}, {Taylor} et~al.}}]{snana}
\bibinfo{author}{\bibfnamefont{R.}~\bibnamefont{{Kessler}}},
  \bibinfo{author}{\bibfnamefont{J.~P.} \bibnamefont{{Bernstein}}},
  \bibinfo{author}{\bibfnamefont{D.}~\bibnamefont{{Cinabro}}},
  \bibinfo{author}{\bibfnamefont{B.}~\bibnamefont{{Dilday}}},
  \bibinfo{author}{\bibfnamefont{J.~A.} \bibnamefont{{Frieman}}},
  \bibinfo{author}{\bibfnamefont{S.}~\bibnamefont{{Jha}}},
  \bibinfo{author}{\bibfnamefont{S.}~\bibnamefont{{Kuhlmann}}},
  \bibinfo{author}{\bibfnamefont{G.}~\bibnamefont{{Miknaitis}}},
  \bibinfo{author}{\bibfnamefont{M.}~\bibnamefont{{Sako}}},
  \bibinfo{author}{\bibfnamefont{M.}~\bibnamefont{{Taylor}}},
  \bibnamefont{et~al.}, \bibinfo{journal}{"Publ. Astron. Soc. Pac."}
  \textbf{\bibinfo{volume}{121}}, \bibinfo{pages}{1028} (\bibinfo{year}{2009}),
  \eprint{0908.4280}.

\bibitem[{\citenamefont{{Hinton} and {Brout}}(2020)}]{pippin}
\bibinfo{author}{\bibfnamefont{S.}~\bibnamefont{{Hinton}}} \bibnamefont{and}
  \bibinfo{author}{\bibfnamefont{D.}~\bibnamefont{{Brout}}},
  \bibinfo{journal}{The Journal of Open Source Software}
  \textbf{\bibinfo{volume}{5}}, \bibinfo{eid}{2122} (\bibinfo{year}{2020}).

\bibitem[{\citenamefont{{Cahn}}(2009)}]{Cahn2009}
\bibinfo{author}{\bibfnamefont{R.~N.} \bibnamefont{{Cahn}}},
  \emph{\bibinfo{title}{{Dark Energy Task Force}}} (\bibinfo{publisher}{World
  Scientific}, \bibinfo{year}{2009}), pp. \bibinfo{pages}{685--695}.

\bibitem[{\citenamefont{Ivezi\'c et~al.}(2019)}]{ivezic}
\bibinfo{author}{\bibfnamefont{v.}~\bibnamefont{Ivezi\'c}} \bibnamefont{et~al.}
  (\bibinfo{collaboration}{LSST}), \bibinfo{journal}{Astrophys. J.}
  \textbf{\bibinfo{volume}{873}}, \bibinfo{pages}{111} (\bibinfo{year}{2019}),
  \eprint{0805.2366}.

\bibitem[{\citenamefont{{LSST Dark Energy Science Collaboration (LSST DESC)}
  et~al.}(2021)\citenamefont{{LSST Dark Energy Science Collaboration (LSST
  DESC)}, {Abolfathi}, {Alonso}, {Armstrong}, {Aubourg}, {Awan}, {Babuji},
  {Bauer}, {Bean}, {Beckett} et~al.}}]{dc2}
\bibinfo{author}{\bibnamefont{{LSST Dark Energy Science Collaboration (LSST
  DESC)}}}, \bibinfo{author}{\bibfnamefont{B.}~\bibnamefont{{Abolfathi}}},
  \bibinfo{author}{\bibfnamefont{D.}~\bibnamefont{{Alonso}}},
  \bibinfo{author}{\bibfnamefont{R.}~\bibnamefont{{Armstrong}}},
  \bibinfo{author}{\bibfnamefont{{\'E}.}~\bibnamefont{{Aubourg}}},
  \bibinfo{author}{\bibfnamefont{H.}~\bibnamefont{{Awan}}},
  \bibinfo{author}{\bibfnamefont{Y.~N.} \bibnamefont{{Babuji}}},
  \bibinfo{author}{\bibfnamefont{F.~E.} \bibnamefont{{Bauer}}},
  \bibinfo{author}{\bibfnamefont{R.}~\bibnamefont{{Bean}}},
  \bibinfo{author}{\bibfnamefont{G.}~\bibnamefont{{Beckett}}},
  \bibnamefont{et~al.}, \bibinfo{journal}{Astrophys. J., Suppl. Ser.}
  \textbf{\bibinfo{volume}{253}}, \bibinfo{eid}{31} (\bibinfo{year}{2021}),
  \eprint{2010.05926}.

\bibitem[{\citenamefont{{S{\'a}nchez} et~al.}(2021)\citenamefont{{S{\'a}nchez},
  {Kessler}, {Scolnic}, {Armstrong}, {Biswas}, {Bogart}, {Chiang},
  {Cohen-Tanugi}, {Fouchez}, {Gris} et~al.}}]{Sanchez}
\bibinfo{author}{\bibfnamefont{B.}~\bibnamefont{{S{\'a}nchez}}},
  \bibinfo{author}{\bibfnamefont{R.}~\bibnamefont{{Kessler}}},
  \bibinfo{author}{\bibfnamefont{D.}~\bibnamefont{{Scolnic}}},
  \bibinfo{author}{\bibfnamefont{B.}~\bibnamefont{{Armstrong}}},
  \bibinfo{author}{\bibfnamefont{R.}~\bibnamefont{{Biswas}}},
  \bibinfo{author}{\bibfnamefont{J.}~\bibnamefont{{Bogart}}},
  \bibinfo{author}{\bibfnamefont{J.}~\bibnamefont{{Chiang}}},
  \bibinfo{author}{\bibfnamefont{J.}~\bibnamefont{{Cohen-Tanugi}}},
  \bibinfo{author}{\bibfnamefont{D.}~\bibnamefont{{Fouchez}}},
  \bibinfo{author}{\bibfnamefont{P.}~\bibnamefont{{Gris}}},
  \bibnamefont{et~al.}, \bibinfo{journal}{arXiv e-prints}
  \bibinfo{eid}{arXiv:2111.06858} (\bibinfo{year}{2021}), \eprint{2111.06858}.

\bibitem[{\citenamefont{Lokken et~al.}(2022)}]{lokken}
\bibinfo{author}{\bibfnamefont{M.}~\bibnamefont{Lokken}} \bibnamefont{et~al.}
  (\bibinfo{collaboration}{LSST Dark Energy Science}) (\bibinfo{year}{2022}),
  \eprint{2206.02815}.

\bibitem[{\citenamefont{{Alard} and {Lupton}}(1998)}]{DIA1}
\bibinfo{author}{\bibfnamefont{C.}~\bibnamefont{{Alard}}} \bibnamefont{and}
  \bibinfo{author}{\bibfnamefont{R.~H.} \bibnamefont{{Lupton}}},
  \bibinfo{journal}{Astrophys. J.} \textbf{\bibinfo{volume}{503}},
  \bibinfo{pages}{325} (\bibinfo{year}{1998}), \eprint{astro-ph/9712287}.

\bibitem[{\citenamefont{{Fukugita} et~al.}(1996)\citenamefont{{Fukugita},
  {Ichikawa}, {Gunn}, {Doi}, {Shimasaku}, and {Schneider}}}]{fukugita}
\bibinfo{author}{\bibfnamefont{M.}~\bibnamefont{{Fukugita}}},
  \bibinfo{author}{\bibfnamefont{T.}~\bibnamefont{{Ichikawa}}},
  \bibinfo{author}{\bibfnamefont{J.~E.} \bibnamefont{{Gunn}}},
  \bibinfo{author}{\bibfnamefont{M.}~\bibnamefont{{Doi}}},
  \bibinfo{author}{\bibfnamefont{K.}~\bibnamefont{{Shimasaku}}},
  \bibnamefont{and} \bibinfo{author}{\bibfnamefont{D.~P.}
  \bibnamefont{{Schneider}}}, \bibinfo{journal}{Astron. J}
  \textbf{\bibinfo{volume}{111}}, \bibinfo{pages}{1748} (\bibinfo{year}{1996}).

\bibitem[{\citenamefont{{Hlo{\v{z}}ek}
  et~al.}(2020)\citenamefont{{Hlo{\v{z}}ek}, {Ponder}, {Malz}, {Dai},
  {Narayan}, {Ishida}, {Allam}, {Bahmanyar}, {Biswas}, {Galbany}
  et~al.}}]{plasticc_H2020}
\bibinfo{author}{\bibfnamefont{R.}~\bibnamefont{{Hlo{\v{z}}ek}}},
  \bibinfo{author}{\bibfnamefont{K.~A.} \bibnamefont{{Ponder}}},
  \bibinfo{author}{\bibfnamefont{A.~I.} \bibnamefont{{Malz}}},
  \bibinfo{author}{\bibfnamefont{M.}~\bibnamefont{{Dai}}},
  \bibinfo{author}{\bibfnamefont{G.}~\bibnamefont{{Narayan}}},
  \bibinfo{author}{\bibfnamefont{E.~E.~O.} \bibnamefont{{Ishida}}},
  \bibinfo{author}{\bibfnamefont{J.}~\bibnamefont{{Allam}}, \bibfnamefont{T.}},
  \bibinfo{author}{\bibfnamefont{A.}~\bibnamefont{{Bahmanyar}}},
  \bibinfo{author}{\bibfnamefont{R.}~\bibnamefont{{Biswas}}},
  \bibinfo{author}{\bibfnamefont{L.}~\bibnamefont{{Galbany}}},
  \bibnamefont{et~al.}, \bibinfo{journal}{arXiv e-prints}
  \bibinfo{eid}{arXiv:2012.12392} (\bibinfo{year}{2020}), \eprint{2012.12392}.

\bibitem[{\citenamefont{{Graham} et~al.}(2018)\citenamefont{{Graham},
  {Connolly}, {Ivezi{\'c}}, {Schmidt}, {Jones}, {Juri{\'c}}, {Daniel}, and
  {Yoachim}}}]{Graham2018_photoz}
\bibinfo{author}{\bibfnamefont{M.~L.} \bibnamefont{{Graham}}},
  \bibinfo{author}{\bibfnamefont{A.~J.} \bibnamefont{{Connolly}}},
  \bibinfo{author}{\bibfnamefont{{\v{Z}}.}~\bibnamefont{{Ivezi{\'c}}}},
  \bibinfo{author}{\bibfnamefont{S.~J.} \bibnamefont{{Schmidt}}},
  \bibinfo{author}{\bibfnamefont{R.~L.} \bibnamefont{{Jones}}},
  \bibinfo{author}{\bibfnamefont{M.}~\bibnamefont{{Juri{\'c}}}},
  \bibinfo{author}{\bibfnamefont{S.~F.} \bibnamefont{{Daniel}}},
  \bibnamefont{and}
  \bibinfo{author}{\bibfnamefont{P.}~\bibnamefont{{Yoachim}}},
  \bibinfo{journal}{Astron. J.} \textbf{\bibinfo{volume}{155}},
  \bibinfo{eid}{1} (\bibinfo{year}{2018}), \eprint{1706.09507}.

\bibitem[{\citenamefont{{Kessler}
  et~al.}(2019{\natexlab{b}})\citenamefont{{Kessler}, {Brout}, {D'Andrea},
  {Davis}, {Hinton}, {Kim}, {Lasker}, {Lidman}, {Macaulay}, {M{\"o}ller}
  et~al.}}]{kessler2019}
\bibinfo{author}{\bibfnamefont{R.}~\bibnamefont{{Kessler}}},
  \bibinfo{author}{\bibfnamefont{D.}~\bibnamefont{{Brout}}},
  \bibinfo{author}{\bibfnamefont{C.~B.} \bibnamefont{{D'Andrea}}},
  \bibinfo{author}{\bibfnamefont{T.~M.} \bibnamefont{{Davis}}},
  \bibinfo{author}{\bibfnamefont{S.~R.} \bibnamefont{{Hinton}}},
  \bibinfo{author}{\bibfnamefont{A.~G.} \bibnamefont{{Kim}}},
  \bibinfo{author}{\bibfnamefont{J.}~\bibnamefont{{Lasker}}},
  \bibinfo{author}{\bibfnamefont{C.}~\bibnamefont{{Lidman}}},
  \bibinfo{author}{\bibfnamefont{E.}~\bibnamefont{{Macaulay}}},
  \bibinfo{author}{\bibfnamefont{A.}~\bibnamefont{{M{\"o}ller}}},
  \bibnamefont{et~al.}, \bibinfo{journal}{Mon. Not. Roy. Astron. Soc.}
  \textbf{\bibinfo{volume}{485}}, \bibinfo{pages}{1171}
  (\bibinfo{year}{2019}{\natexlab{b}}), \eprint{1811.02379}.

\bibitem[{\citenamefont{{Scolnic} and {Kessler}}(2016)}]{SK2016}
\bibinfo{author}{\bibfnamefont{D.}~\bibnamefont{{Scolnic}}} \bibnamefont{and}
  \bibinfo{author}{\bibfnamefont{R.}~\bibnamefont{{Kessler}}},
  \bibinfo{journal}{Astrophys. J. Lett.} \textbf{\bibinfo{volume}{822}},
  \bibinfo{eid}{L35} (\bibinfo{year}{2016}), \eprint{1603.01559}.

\bibitem[{\citenamefont{{Pierel} et~al.}(2018)\citenamefont{{Pierel}, {Rodney},
  {Avelino}, {Bianco}, {Filippenko}, {Foley}, {Friedman}, {Hicken}, {Hounsell},
  {Jha} et~al.}}]{salt+}
\bibinfo{author}{\bibfnamefont{J.~D.~R.} \bibnamefont{{Pierel}}},
  \bibinfo{author}{\bibfnamefont{S.}~\bibnamefont{{Rodney}}},
  \bibinfo{author}{\bibfnamefont{A.}~\bibnamefont{{Avelino}}},
  \bibinfo{author}{\bibfnamefont{F.}~\bibnamefont{{Bianco}}},
  \bibinfo{author}{\bibfnamefont{A.~V.} \bibnamefont{{Filippenko}}},
  \bibinfo{author}{\bibfnamefont{R.~J.} \bibnamefont{{Foley}}},
  \bibinfo{author}{\bibfnamefont{A.}~\bibnamefont{{Friedman}}},
  \bibinfo{author}{\bibfnamefont{M.}~\bibnamefont{{Hicken}}},
  \bibinfo{author}{\bibfnamefont{R.}~\bibnamefont{{Hounsell}}},
  \bibinfo{author}{\bibfnamefont{S.~W.} \bibnamefont{{Jha}}},
  \bibnamefont{et~al.}, \bibinfo{journal}{Publ. Astron. Soc. Pac.}
  \textbf{\bibinfo{volume}{130}}, \bibinfo{pages}{114504}
  (\bibinfo{year}{2018}), \eprint{1808.02534}.

\bibitem[{\citenamefont{{Schlafly} and {Finkbeiner}}(2011)}]{ebv2}
\bibinfo{author}{\bibfnamefont{E.~F.} \bibnamefont{{Schlafly}}}
  \bibnamefont{and} \bibinfo{author}{\bibfnamefont{D.~P.}
  \bibnamefont{{Finkbeiner}}}, \bibinfo{journal}{Astrophys. J.}
  \textbf{\bibinfo{volume}{737}}, \bibinfo{eid}{103} (\bibinfo{year}{2011}),
  \eprint{1012.4804}.

\bibitem[{\citenamefont{{Delgado} et~al.}(2014)\citenamefont{{Delgado}, {Saha},
  {Chandrasekharan}, {Cook}, {Petry}, and {Ridgway}}}]{OpSim1}
\bibinfo{author}{\bibfnamefont{F.}~\bibnamefont{{Delgado}}},
  \bibinfo{author}{\bibfnamefont{A.}~\bibnamefont{{Saha}}},
  \bibinfo{author}{\bibfnamefont{S.}~\bibnamefont{{Chandrasekharan}}},
  \bibinfo{author}{\bibfnamefont{K.}~\bibnamefont{{Cook}}},
  \bibinfo{author}{\bibfnamefont{C.}~\bibnamefont{{Petry}}}, \bibnamefont{and}
  \bibinfo{author}{\bibfnamefont{S.}~\bibnamefont{{Ridgway}}}, in
  \emph{\bibinfo{booktitle}{Modeling, Systems Engineering, and Project
  Management \- for Astronomy VI}}, edited by
  \bibinfo{editor}{\bibfnamefont{G.~Z.} \bibnamefont{{Angeli}}}
  \bibnamefont{and}
  \bibinfo{editor}{\bibfnamefont{P.}~\bibnamefont{{Dierickx}}}
  (\bibinfo{year}{2014}), vol. \bibinfo{volume}{9150} of
  \emph{\bibinfo{series}{Society of Photo-Optical Instrumentation Engineers}},
  p. \bibinfo{pages}{915015}.

\bibitem[{OpS(2016)}]{OpSim2}
\emph{\bibinfo{title}{{Observatory Operations: Strategies, Processes, and
  Systems VI}}}, vol. \bibinfo{volume}{9910} of \emph{\bibinfo{series}{Society
  of Photo-Optical Instrumentation Engineers}} (\bibinfo{year}{2016}).

\bibitem[{\citenamefont{{Reuter} et~al.}(2016)}]{OpSim3}
\bibinfo{author}{\bibfnamefont{M.~A.} \bibnamefont{{Reuter}}}
  \bibnamefont{et~al.}, in \emph{\bibinfo{booktitle}{Modeling, Systems
  Engineering, and Project Management \- for Astronomy VI}}, edited by
  \bibinfo{editor}{\bibfnamefont{G.~Z.} \bibnamefont{{Angeli}}}
  \bibnamefont{and}
  \bibinfo{editor}{\bibfnamefont{P.}~\bibnamefont{{Dierickx}}}
  (\bibinfo{year}{2016}), vol. \bibinfo{volume}{9911} of
  \emph{\bibinfo{series}{Society of Photo-Optical Instrumentation Engineers
  (SPIE)}}, p. \bibinfo{pages}{991125}.

\bibitem[{\citenamefont{{de Jong} et~al.}(2019)\citenamefont{{de Jong},
  {Agertz}, {Berbel}, {Aird}, {Alexander} et~al.}}]{4MOST2}
\bibinfo{author}{\bibfnamefont{R.~S.} \bibnamefont{{de Jong}}},
  \bibinfo{author}{\bibfnamefont{O.}~\bibnamefont{{Agertz}}},
  \bibinfo{author}{\bibfnamefont{A.~A.} \bibnamefont{{Berbel}}},
  \bibinfo{author}{\bibfnamefont{J.}~\bibnamefont{{Aird}}},
  \bibinfo{author}{\bibfnamefont{D.~A.} \bibnamefont{{Alexander}}},
  \bibnamefont{et~al.}, \bibinfo{journal}{The Messenger}
  \textbf{\bibinfo{volume}{175}}, \bibinfo{pages}{3} (\bibinfo{year}{2019}),
  \eprint{1903.02464}.

\bibitem[{\citenamefont{{Dilday} et~al.}(2008)\citenamefont{{Dilday},
  {Kessler}, {Frieman}, {Holtzman}, {Marriner}, {Miknaitis}, {Nichol},
  {Romani}, {Sako}, {Bassett} et~al.}}]{Dilday2008}
\bibinfo{author}{\bibfnamefont{B.}~\bibnamefont{{Dilday}}},
  \bibinfo{author}{\bibfnamefont{R.}~\bibnamefont{{Kessler}}},
  \bibinfo{author}{\bibfnamefont{J.~A.} \bibnamefont{{Frieman}}},
  \bibinfo{author}{\bibfnamefont{J.}~\bibnamefont{{Holtzman}}},
  \bibinfo{author}{\bibfnamefont{J.}~\bibnamefont{{Marriner}}},
  \bibinfo{author}{\bibfnamefont{G.}~\bibnamefont{{Miknaitis}}},
  \bibinfo{author}{\bibfnamefont{R.~C.} \bibnamefont{{Nichol}}},
  \bibinfo{author}{\bibfnamefont{R.}~\bibnamefont{{Romani}}},
  \bibinfo{author}{\bibfnamefont{M.}~\bibnamefont{{Sako}}},
  \bibinfo{author}{\bibfnamefont{B.}~\bibnamefont{{Bassett}}},
  \bibnamefont{et~al.}, \bibinfo{journal}{Astrophys. J.}
  \textbf{\bibinfo{volume}{682}}, \bibinfo{pages}{262} (\bibinfo{year}{2008}),
  \eprint{0801.3297}.

\bibitem[{\citenamefont{{Hounsell} et~al.}(2018)\citenamefont{{Hounsell},
  {Scolnic}, {Foley}, {Kessler}, {Miranda}, {Avelino}, {Bohlin}, {Filippenko},
  {Frieman}, {Jha} et~al.}}]{Hounsell2018}
\bibinfo{author}{\bibfnamefont{R.}~\bibnamefont{{Hounsell}}},
  \bibinfo{author}{\bibfnamefont{D.}~\bibnamefont{{Scolnic}}},
  \bibinfo{author}{\bibfnamefont{R.~J.} \bibnamefont{{Foley}}},
  \bibinfo{author}{\bibfnamefont{R.}~\bibnamefont{{Kessler}}},
  \bibinfo{author}{\bibfnamefont{V.}~\bibnamefont{{Miranda}}},
  \bibinfo{author}{\bibfnamefont{A.}~\bibnamefont{{Avelino}}},
  \bibinfo{author}{\bibfnamefont{R.~C.} \bibnamefont{{Bohlin}}},
  \bibinfo{author}{\bibfnamefont{A.~V.} \bibnamefont{{Filippenko}}},
  \bibinfo{author}{\bibfnamefont{J.}~\bibnamefont{{Frieman}}},
  \bibinfo{author}{\bibfnamefont{S.~W.} \bibnamefont{{Jha}}},
  \bibnamefont{et~al.}, \bibinfo{journal}{Astrophys. J.}
  \textbf{\bibinfo{volume}{867}}, \bibinfo{eid}{23} (\bibinfo{year}{2018}),
  \eprint{1702.01747}.

\bibitem[{\citenamefont{{Guy} et~al.}(2010)}]{Guy2010}
\bibinfo{author}{\bibfnamefont{J.}~\bibnamefont{{Guy}}} \bibnamefont{et~al.},
  \bibinfo{journal}{Astron. Astrophys.} \textbf{\bibinfo{volume}{523}},
  \bibinfo{eid}{A7} (\bibinfo{year}{2010}), \eprint{1010.4743}.

\bibitem[{\citenamefont{{Conley} et~al.}(2011)\citenamefont{{Conley}, {Guy},
  {Sullivan}, {Regnault}, {Astier}, {Balland}, {Basa}, {Carlberg}, {Fouchez},
  {Hardin} et~al.}}]{conley11}
\bibinfo{author}{\bibfnamefont{A.}~\bibnamefont{{Conley}}},
  \bibinfo{author}{\bibfnamefont{J.}~\bibnamefont{{Guy}}},
  \bibinfo{author}{\bibfnamefont{M.}~\bibnamefont{{Sullivan}}},
  \bibinfo{author}{\bibfnamefont{N.}~\bibnamefont{{Regnault}}},
  \bibinfo{author}{\bibfnamefont{P.}~\bibnamefont{{Astier}}},
  \bibinfo{author}{\bibfnamefont{C.}~\bibnamefont{{Balland}}},
  \bibinfo{author}{\bibfnamefont{S.}~\bibnamefont{{Basa}}},
  \bibinfo{author}{\bibfnamefont{R.~G.} \bibnamefont{{Carlberg}}},
  \bibinfo{author}{\bibfnamefont{D.}~\bibnamefont{{Fouchez}}},
  \bibinfo{author}{\bibfnamefont{D.}~\bibnamefont{{Hardin}}},
  \bibnamefont{et~al.}, \bibinfo{journal}{Astrophys. J. Lett. Suppl.}
  \textbf{\bibinfo{volume}{192}}, \bibinfo{eid}{1} (\bibinfo{year}{2011}),
  \eprint{1104.1443}.

\bibitem[{\citenamefont{Brout et~al.}(2019)}]{Brout2019_DES3YR_ANA}
\bibinfo{author}{\bibfnamefont{D.}~\bibnamefont{Brout}} \bibnamefont{et~al.}
  (\bibinfo{collaboration}{DES}), \bibinfo{journal}{Astrophys. J.}
  \textbf{\bibinfo{volume}{874}}, \bibinfo{pages}{150} (\bibinfo{year}{2019}),
  \eprint{1811.02377}.

\bibitem[{\citenamefont{{Bohlin} et~al.}(2014)\citenamefont{{Bohlin}, {Gordon},
  and {Tremblay}}}]{bohlin:calibrationHST}
\bibinfo{author}{\bibfnamefont{R.~C.} \bibnamefont{{Bohlin}}},
  \bibinfo{author}{\bibfnamefont{K.~D.} \bibnamefont{{Gordon}}},
  \bibnamefont{and} \bibinfo{author}{\bibfnamefont{P.~E.}
  \bibnamefont{{Tremblay}}}, \bibinfo{journal}{Publ. Astron. Soc. Pac.}
  \textbf{\bibinfo{volume}{126}}, \bibinfo{pages}{711} (\bibinfo{year}{2014}),
  \eprint{1406.1707}.

\bibitem[{\citenamefont{Ivezi{\'c} et~al.}(2018)}]{LSST_scibook}
\bibinfo{author}{\bibfnamefont{{\v{Z}}.}~\bibnamefont{Ivezi{\'c}}}
  \bibnamefont{et~al.} (\bibinfo{year}{2018}).

\bibitem[{\citenamefont{{Schlafly} et~al.}(2012)\citenamefont{{Schlafly},
  {Finkbeiner}, {Juri{\'c}}, {Magnier}, {Burgett}, {Chambers}, {Grav},
  {Hodapp}, {Kaiser}, {Kudritzki} et~al.}}]{Schlafly2012}
\bibinfo{author}{\bibfnamefont{E.~F.} \bibnamefont{{Schlafly}}},
  \bibinfo{author}{\bibfnamefont{D.~P.} \bibnamefont{{Finkbeiner}}},
  \bibinfo{author}{\bibfnamefont{M.}~\bibnamefont{{Juri{\'c}}}},
  \bibinfo{author}{\bibfnamefont{E.~A.} \bibnamefont{{Magnier}}},
  \bibinfo{author}{\bibfnamefont{W.~S.} \bibnamefont{{Burgett}}},
  \bibinfo{author}{\bibfnamefont{K.~C.} \bibnamefont{{Chambers}}},
  \bibinfo{author}{\bibfnamefont{T.}~\bibnamefont{{Grav}}},
  \bibinfo{author}{\bibfnamefont{K.~W.} \bibnamefont{{Hodapp}}},
  \bibinfo{author}{\bibfnamefont{N.}~\bibnamefont{{Kaiser}}},
  \bibinfo{author}{\bibfnamefont{R.~P.} \bibnamefont{{Kudritzki}}},
  \bibnamefont{et~al.}, \bibinfo{journal}{Astrophys. J.}
  \textbf{\bibinfo{volume}{756}}, \bibinfo{eid}{158} (\bibinfo{year}{2012}),
  \eprint{1201.2208}.

\bibitem[{\citenamefont{{Magnier} et~al.}(2013)\citenamefont{{Magnier},
  {Schlafly}, {Finkbeiner}, {Juric}, {Tonry}, {Burgett}, {Chambers},
  {Flewelling}, {Kaiser}, {Kudritzki} et~al.}}]{PS1cal2013}
\bibinfo{author}{\bibfnamefont{E.~A.} \bibnamefont{{Magnier}}},
  \bibinfo{author}{\bibfnamefont{E.}~\bibnamefont{{Schlafly}}},
  \bibinfo{author}{\bibfnamefont{D.}~\bibnamefont{{Finkbeiner}}},
  \bibinfo{author}{\bibfnamefont{M.}~\bibnamefont{{Juric}}},
  \bibinfo{author}{\bibfnamefont{J.~L.} \bibnamefont{{Tonry}}},
  \bibinfo{author}{\bibfnamefont{W.~S.} \bibnamefont{{Burgett}}},
  \bibinfo{author}{\bibfnamefont{K.~C.} \bibnamefont{{Chambers}}},
  \bibinfo{author}{\bibfnamefont{H.~A.} \bibnamefont{{Flewelling}}},
  \bibinfo{author}{\bibfnamefont{N.}~\bibnamefont{{Kaiser}}},
  \bibinfo{author}{\bibfnamefont{R.~P.} \bibnamefont{{Kudritzki}}},
  \bibnamefont{et~al.}, \bibinfo{journal}{Astrophys. J. Lett. Suppl.}
  \textbf{\bibinfo{volume}{205}}, \bibinfo{eid}{20} (\bibinfo{year}{2013}),
  \eprint{1303.3634}.

\bibitem[{\citenamefont{{Calcino} and {Davis}}(2017)}]{Calcino2017}
\bibinfo{author}{\bibfnamefont{J.}~\bibnamefont{{Calcino}}} \bibnamefont{and}
  \bibinfo{author}{\bibfnamefont{T.}~\bibnamefont{{Davis}}},
  \bibinfo{journal}{J. Cosmol. Astropart. Phys.}
  \textbf{\bibinfo{volume}{2017}}, \bibinfo{eid}{038} (\bibinfo{year}{2017}),
  \eprint{1610.07695}.

\bibitem[{\citenamefont{{Myles} et~al.}(2021)\citenamefont{{Myles}, {Alarcon},
  {Amon}, {S{\'a}nchez}, {Everett}, {DeRose}, {McCullough}, {Gruen},
  {Bernstein}, {Troxel} et~al.}}]{PZ2021_DES3YR_WL}
\bibinfo{author}{\bibfnamefont{J.}~\bibnamefont{{Myles}}},
  \bibinfo{author}{\bibfnamefont{A.}~\bibnamefont{{Alarcon}}},
  \bibinfo{author}{\bibfnamefont{A.}~\bibnamefont{{Amon}}},
  \bibinfo{author}{\bibfnamefont{C.}~\bibnamefont{{S{\'a}nchez}}},
  \bibinfo{author}{\bibfnamefont{S.}~\bibnamefont{{Everett}}},
  \bibinfo{author}{\bibfnamefont{J.}~\bibnamefont{{DeRose}}},
  \bibinfo{author}{\bibfnamefont{J.}~\bibnamefont{{McCullough}}},
  \bibinfo{author}{\bibfnamefont{D.}~\bibnamefont{{Gruen}}},
  \bibinfo{author}{\bibfnamefont{G.~M.} \bibnamefont{{Bernstein}}},
  \bibinfo{author}{\bibfnamefont{M.~A.} \bibnamefont{{Troxel}}},
  \bibnamefont{et~al.}, \bibinfo{journal}{Mon. Not. Roy. Astron. Soc.}
  \textbf{\bibinfo{volume}{505}}, \bibinfo{pages}{4249} (\bibinfo{year}{2021}),
  \eprint{2012.08566}.

\bibitem[{\citenamefont{{Tripp}}(1998)}]{Trip1998}
\bibinfo{author}{\bibfnamefont{R.}~\bibnamefont{{Tripp}}},
  \bibinfo{journal}{Astron. Astrophys.} \textbf{\bibinfo{volume}{331}},
  \bibinfo{pages}{815} (\bibinfo{year}{1998}).

\bibitem[{\citenamefont{{Komatsu} et~al.}(2009)}]{komatsu}
\bibinfo{author}{\bibfnamefont{E.}~\bibnamefont{{Komatsu}}}
  \bibnamefont{et~al.}, \bibinfo{journal}{Astrophys. J.s}
  \textbf{\bibinfo{volume}{180}}, \bibinfo{pages}{330} (\bibinfo{year}{2009}),
  \eprint{0803.0547}.

\bibitem[{\citenamefont{{Planck Collaboration}
  et~al.}(2020)\citenamefont{{Planck Collaboration}, {Aghanim}, {Akrami},
  {Ashdown}, {Aumont}, {Baccigalupi} et~al.}}]{Planck2018}
\bibinfo{author}{\bibnamefont{{Planck Collaboration}}},
  \bibinfo{author}{\bibfnamefont{N.}~\bibnamefont{{Aghanim}}},
  \bibinfo{author}{\bibfnamefont{Y.}~\bibnamefont{{Akrami}}},
  \bibinfo{author}{\bibfnamefont{M.}~\bibnamefont{{Ashdown}}},
  \bibinfo{author}{\bibfnamefont{J.}~\bibnamefont{{Aumont}}},
  \bibinfo{author}{\bibfnamefont{C.}~\bibnamefont{{Baccigalupi}}},
  \bibnamefont{et~al.} (\bibinfo{collaboration}{Planck}),
  \bibinfo{journal}{Astron. Astrophys.} \textbf{\bibinfo{volume}{641}},
  \bibinfo{pages}{A6} (\bibinfo{year}{2020}), \bibinfo{note}{[Erratum:
  Astron.Astrophys. 652, C4 (2021)]}, \eprint{1807.06209}.

\bibitem[{\citenamefont{Albrecht et~al.}(2006)}]{FoM}
\bibinfo{author}{\bibfnamefont{A.}~\bibnamefont{Albrecht}}
  \bibnamefont{et~al.}, \bibinfo{journal}{arXiv e-prints}
  (\bibinfo{year}{2006}), \eprint{astro-ph/0609591}.

\bibitem[{\citenamefont{{Brout} et~al.}(2021)\citenamefont{{Brout}, {Hinton},
  and {Scolnic}}}]{Binning_is_sinning}
\bibinfo{author}{\bibfnamefont{D.}~\bibnamefont{{Brout}}},
  \bibinfo{author}{\bibfnamefont{S.~R.} \bibnamefont{{Hinton}}},
  \bibnamefont{and}
  \bibinfo{author}{\bibfnamefont{D.}~\bibnamefont{{Scolnic}}},
  \bibinfo{journal}{Astrophysical Journal, Letters}
  \textbf{\bibinfo{volume}{912}}, \bibinfo{eid}{L26} (\bibinfo{year}{2021}),
  \eprint{2012.05900}.

\end{thebibliography}
\end{document}